\documentclass[preprint,aps]{revtex4}
\usepackage{graphicx}
\usepackage{booktabs}
\usepackage{multirow}
\usepackage{setspace}
\usepackage{amsmath}
\usepackage{cancel}
\usepackage{rotating}
\usepackage{mathtools}
\usepackage{endnotes}
\usepackage{siunitx}
\let\footnote=\endnote

\usepackage[usenames, dvipsnames]{color}
\usepackage[ pdftex, plainpages = false, pdfpagelabels, 
                 pdfpagelayout = useoutlines,
                 bookmarks,
                 bookmarksopen = true,
                 bookmarksnumbered = true,
                 breaklinks = true,
                 linktocpage=all,
                 pagebackref=false,
                 colorlinks = true,
                 linkcolor = BrickRed,
                 urlcolor  = blue,
                 citecolor = BrickRed,
                 anchorcolor = green,
                 hyperindex = true,
                 hyperfigures
                 ]{hyperref} 

\newcommand{\comment}[1]{}

\begin{document}
\title{\href{http://necsi.edu/research/economics/econuniversal}{Preliminary steps toward a universal economic dynamics for monetary and fiscal policy}} 
\author{
Yaneer Bar-Yam, Jean Langlois-Meurinne, Mari Kawakatsu, Rodolfo Garcia}   
\affiliation{New England Complex Systems Institute,
277 Broadway, Cambridge, MA 02139, USA}
\date{\today}

\begin{abstract}
We consider the relationship between economic activity and intervention, including monetary and fiscal policy, using a universal monetary and response dynamics framework. Central bank policies are designed for economic growth without excess inflation. However, unemployment, investment, consumption, and inflation are interlinked. Understanding dynamics is crucial to assessing the effects of policy, especially in the aftermath of the recent financial crisis. Here we lay out a program of research into monetary and economic dynamics and preliminary steps toward its execution. We use general principles of response theory to derive specific implications for policy. We find that the current approach, which considers the overall supply of money to the economy, is insufficient to effectively regulate economic growth. While it can achieve some degree of control, optimizing growth also requires a fiscal policy balancing monetary injection between two dominant loop flows, the consumption and wages loop, and investment and returns loop. The balance arises from a composite of government tax, entitlement, subsidy policies, corporate policies, as well as monetary policy. We further show that empirical evidence is consistent with a transition in 1980 between two regimes---from an oversupply to the consumption and wages loop, to an oversupply of the investment and returns loop. The imbalance is manifest in savings and borrowing by consumers and investors, and in inflation. The latter followed an increasing trend until 1980, and a decreasing one since then, resulting in a zero interest rate largely unrelated to the financial crisis. Three recessions and the financial crisis are part of this dynamic. Optimizing growth now requires shifting the balance. Our analysis supports advocates of greater income and / or government support for the poor who use a larger fraction of income for consumption. This promotes investment due to the growth in expenditures. Otherwise, investment has limited opportunities to gain returns above inflation so capital remains uninvested, and does not contribute to the growth of economic activity.
\end{abstract}

\maketitle

\section{Overview}

Complexity science \cite{baryam1997} provides frameworks that may be helpful in advancing economic theory for better application to real world conditions \cite{arthur1999complexity,Colander2000Complexity,sterman2000business,tesfatsion2003agent,foster2004applied,foster2005simplistic,beinhocker2006origin,holt2011complexity,gallegati2012reconstructing,foxon2012towards,durlauf2012complexity,arthur2013complexity}. The term ``complexity" is often interpreted to mean that models resulting from this approach will be of greater complexity and more difficult to understand. However, one of the key methods of complexity science is to recognize how to simplify models to include only the most relevant parameters through multiscale information and representations \cite{bar2016big,harmon2015anticipating,lagi2015accurate}. This approach seeks to identify the minimal model that captures the macroscopic behavior of a system by aggregating elements to identify their collective behaviors. Minimal models are possible because of ``universality," according to which systems may have the same large scale behavior even if more detailed descriptions are distinct. As a strategy this approach is compatible with macroeconomics that seeks to quantitatively characterize the behavior of the entire economy, as well as the imperative to provide guidance to policies that are concerned with the largest scale of a social system. Simpler models with fewer parameters can be validated more readily against real world data providing confidence that the right principles and concepts are being incorporated in those models. Our objective is to take an exploratory step toward applying such an approach for macroeconomic modeling and monetary policy. 

Current monetary policy adjusts interest rates to regulate inflation and economic activity, particularly employment. There is one regulatory dial, the federal funds interest rate. Regulation consists of adjusting this rate in response to fluctuations of overall economic activity to adjust the activity toward an optimal inflation, or perhaps a tradeoff between inflation and unemployment. This approach assumes a single aggregated measure of economic activity and interest rate adjustments are guided by assuming a linear response to changes in interest rates. In one of the foundational studies of economic growth \cite{Solow01021956} Solow states in discussing liquidity, `But it is exactly here that the futility of trying to describe this situation in terms of a ``real" neoclassical model becomes glaringly evident. Because now one can no longer bypass the direct leverage of monetary factors on real consumption and investment ... there is no dodging the need for a monetary dynamics.'

In performing a preliminary step to develop complexity science based aggregated models, we bypass the formal aggregation machinery and are instead guided by well-established simplified macroeconomic models that aggregate transactions into large scale financial flows. The models on which our analysis is based suggest there is a need to represent distinct flows that are essential for two factors of economic production: labor and capital. As has been well motivated in the economics literature, in order to represent these flows, there is a need to treat consistently the behavior at the same scale, and thus to describe the closure of the flows in self-consistent persistent loops. Thus the models describe economic and monetary dynamics distinguishing two primary monetary loops (1) consumption and wages, and (2) investment and returns (rents). Such a framework is manifest, for example, in the Goodwin model \cite{goodwin1967growth}, and the Kalecki model \cite{kalecki1954theory}, used to describe macroeconomic oscillations called ``business cycles." 

Monetary dynamics has been the subject of increasing interest in macroeconomic modeling. The Goodwin and Kalecki models are examples of a larger class of models that consider the closures of monetary flow and the economic dependencies that constrain variations in those flows. They differ from neoclassical and other models that attribute economic variability directly to shocks without their monetary flow implications. There is a substantial literature addressing general principles as well as numerical models, including consideration of the implications of monetary flows, credit, financial assets and institutions \cite{backus1980model,tobin1969va,tobin1982money,moore1988horizontalists,moore2006shaking,carvalho1992mr}, now flourishing under the name Post Keneysian stock-flow-consistent modeling \cite{godley1999money,bellofiore2000marx,parguez2000credit,lavoie2001endogenous,lavoie2004circuit,lavoie2001kaleckian,gnos200315,graziani2003monetary,seccareccia2003pricing,dos2004post,zezza2004role,zezza2006distribution,godley2006monetary,van2008synthetic,le2008post,le2009financial,dallery2011conflicting,caverzasi2014post}. Various extensions of the Goodwin and Kalecki models have been made that embody specific concepts, address a wide range of limitations of the original models, and compare with data \cite{solow1990goodwin,keen1995finance,keen2001debunking,franke2006wage,barbosa2006distributive,veneziani2006structural,desai2006clarification,miloslav2006goodwin,taylor2012growth,moura2013testing,keen2013monetary,keen2014secular,huu2014orbits,lipton2015modern,tran2016generalized,davila2016goodwin}. Among these, of particular interest for comparing with the present work are those of Steve Keen \cite{keen1995finance} and Lance Taylor \cite{taylor2012growth}. An agent-based model treatment of monetary policy was recently performed by Jean-Philippe Bouchaud et al \cite{bouchaud2017optimal}.

These models use equations that embody assumptions about inflation and production in relation to rates of employment, wages, and investment. It is hard, however, to tell which of many possible assumptions about models are the correct ones and how those assumptions influence policy recommendations that arise from them. Here we focus on reducing the set of assumptions that are present in order to better clarify the role of interventions on economic activity. Motivated by considerations of universality and the relevance of large scale behavior to policy interventions, we consider descriptions designed to capture the primary large scale collective behavior. We expand the equations to first order in small variations of the persistent macroeconomic flows. Since the primary behavior can be described using linear dynamical equations, we consider the universal linear equations sufficient to describe the two primary monetary flows. 

The model has two regimes, an exponential with an oscillation (typically considered in discussing the Goodwin model), and a purely exponential regime. These regimes arise when there are imaginary and real eigenvalues of the relevant characteristic equation, respectively. We determine parameters of the model from data from 1960 to 2014. We obtain reasonable fits by allowing for a regime change in 1980, from exponential to oscillatory, a regime that lasts until the financial crisis. The period since the financial crisis is not long enough to determine the parameters with great confidence, though there may be a second regime change. We then consider more generally the dynamical response of this system to interventions, a response that depends on quadratic terms. While we do not analyze the terms at this order, the principles of dynamic response are sufficient to draw essential conclusions about the existence of two regimes and their behavioral attributes. Comparing the behavioral attributes to observations suggests the transition in 1980 is a regime transition from an oversupply to the wages and consumption cycle prior to 1980 to an oversupply of the investment and returns cycle afterwards. Since then, periodic recessions triggering monetary policy interventions that progressively reduced interest rates from over $10\%$ to $0\%$ have sustained economic development, balancing the monetary flows through accumulation of consumer debt and capital savings. We derive policy implications from these conclusions even without a complete analysis of the underlying dynamical behavior, a task we leave to future work. 

One of the central initial assumptions and ultimately one of the results of our work is the macroeconomic relevance of the two monetary flows present in the Goodwin and Kalecki models. Our analysis provides empirical evidence that these are relevant parameters by showing that recessions occurred for a specific ratio of investment to consumption over a period of 30 years. In every case where investment increased to reach $2/3$ of consumption (in proportion consumption declined to $1.5$ times investment) a recession occurred. This happened in recessions in the early 1980s, 1990s, 2000s, and the 2007 financial crisis.

Our analysis suggests that economic regulation by a single regulatory dial is inadequate. As was stated first by Jan Tinbergen \cite{tinbergen1952theory} and later in the control theory work of Ashby \cite{ashby1958requisite}, there is need for as many regulatory dials as there are dimensions of system variation to regulate. The growth and balance of the two primary monetary flows require not just monetary policy but also fiscal policy. The regime change in 1980, which was likely the result of the Reagan era tax policy changes in the early 1980s, was a necessary step to redirect the economy away from runaway inflation. An equally bold change in policy is needed to stabilize economic growth today. The essence of that policy is redirecting monetary injections toward labor and away from capital due to an imbalance in flows. Conceptually, this is a shift from supply side to demand promoting policies. Absent an increase in demand there is no reason for investors to invest as weak growth in consumer demand means that they will not receive return on their investments. These results are supported by the observation of increasing consumer debt, increasing investor savings, and decreasing interest rates. Perhaps a good analogy is that monetary policy today is like driving a car using only the gas pedal and not using the steering wheel. Having hit the roadside guard rail, we need a change of direction that cannot be sustained through monetary policy alone.

Our analysis provides a framework that is consistent with and can clarify recent discussions of `secular stagnation' \cite{Summers,secularstagnation,Eggertsson,gordon2015secular}. These discussions suggest there are systemic reasons for low growth rates, inferring that the long term ``natural" interest rates that lead to effective economic activity have become negative. Accordingly, economic regulation needs to move beyond a response to economic fluctuations. Our approach identifies a more focused set of policy options because the multiple factors that are considered in those discussions, including demography, technology and inequality, are understood through our analysis to influence economic activity through the relevant monetary flows.

\section{Economic activity in terms of flows} \label{themodel}

\begin{figure}[!tb]  
	\centering	\includegraphics[width=0.8\textwidth]{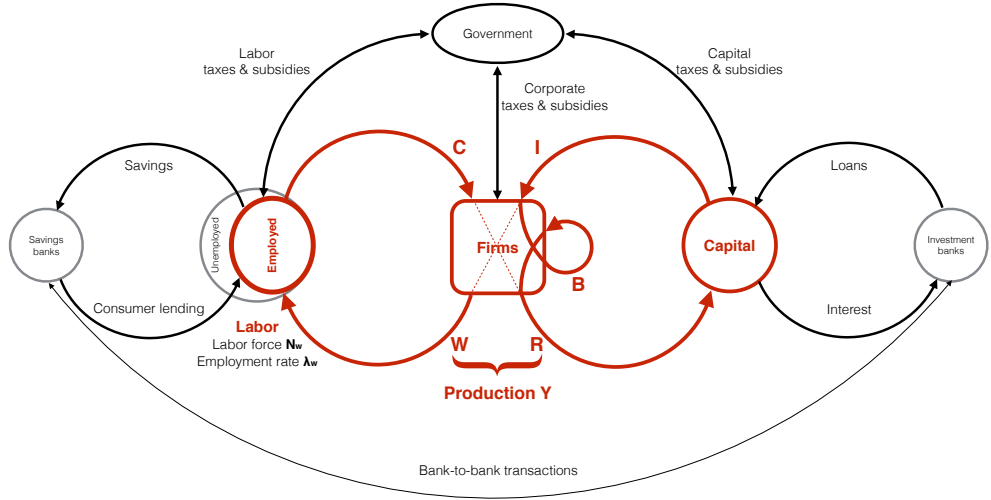}
	\caption{Schematic model of monetary flow representing the wages and consumption loop and capital and return loop (red). Transfers from or to banks (savings and loans) and government (taxes, transfers, subsidies and other economic activities) are also indicated (black).}
	\label{fig:model}
\end{figure}

A simple and standard economic model having two dominant cycles is shown in Fig. \ref{fig:model}. The economy is represented by aggregate agents for labor, capital, firms, banks (investment and savings), and government. Workers receive wages $W$ and consume an amount $C$; capital owners invest an amount $I$ and receive returns (rents) and principal $R$. The possibility of unemployment is indicated graphically though, in describing the flows, the amount of unemployment need not be treated as a separate variable. All firms are represented as one sector. Firm production in monetary terms (GDP) is given by $Y$. Capital expenditures and other inter-firm transactions are $B$. The six variables ($B, Y, W, C, I, R$) are in units of money per time. The consumption and wages cycle and the investment and returns cycle flows are linked through the production of firms that use investment to obtain equipment and buildings that are needed for production. We note that economic activity includes the purchase of such resources so that there is an additional component of production, and the total economic activity is the sum of consumption by labor and inter-firm purchases.

In the simplest framing each of the two cycles must close so that the amount of wages and consumption match, as well as the investment and the total of repayment and returns. This closure is built into the Goodwin model which does not include banks or government. The money supply is included implicitly in the Goodwin model using an inflation rate which increases the economic flows, though the mechanism for adding money to the system is not present in the model (see Appendix \ref{appendix:injection}). In a more complete picture, the labor and capital cycles are augmented by the banking sector (including the central bank) and government (including the treasury) that change the supply of money running through these cycles. In such a more complete picture, inflation would be linked to additional money injected into the cycles as well as changes in the velocity of money. Whether money is printed or borrowed with interest and capital return is relevant. In the latter case, central bank policies, reserve requirements, purchases and interest changes are mechanisms for regulation of monetary injection.

In the model of Fig.  \ref{fig:model}, the banking sector has two branches for labor and capital transactions. While these two branches may interact with each other, traditionally, regulations maintained a degree of separation. Workers can deposit part of their income as savings, or borrow money in mortgages, auto loans, and through credit cards and other consumer debt instruments. Banks give loans to capital owners, who pay back interest in return. We may include loans to corporations in this category, or treat it separately. Finally, the government collects taxes from and provides subsidies to each of labor, capital, and firms, and participates in economic activity. Inclusion of a central bank would add an agent that interacts with banks and government through loans, government debt instruments and payments to the treasury. Inclusion of a treasury agent would separate aspects of government actions. For our current purposes, the specifics of the banking-central bank-government system are not essential. 

As a first step, we analyze a simple economy consisting of labor, capital, and firms only (highlighted in red in Fig. \ref{fig:model}).

\section{Relationship to Goodwin's Predator-Prey Model} \label{goodwin}
We introduce Goodwin's model as a historical motivation for the mathematical strategy we will use later. Most of the assumptions of this model will not be used, and are not relevant to our analysis. Goodwin's predator-prey model \cite{goodwin1967growth} assumes a simplified economy of one type of good. Let $N_w, \lambda_w, \theta_w$, and $P$ be the available labor force measured in number of workers, the employment rate measured as the fraction of the labor force that is employed, the productivity of workers measured in units per worker per unit time, and the price in monetary units, respectively. The number of employed workers is $N_w \lambda_w$.  The firms' production is given by
\begin{equation}
Y = \lambda_w \theta_w N_w P \label{eq:Y}
\end{equation} 
It is assumed that the productivity of labor, labor force, and price grow exponentially at given rates, so that
\begin{align}
\theta_w = e^{\alpha t}, \qquad 
N_w = e^{\beta t}, \qquad
P = e^{\gamma t}, \label{eq:lambda1}
\end{align}
where $\alpha, \beta, \gamma$ are constants. Thus, aside from a pre-specified exponential growth, the dynamics of production is set by employment $\lambda_w$.

In the Goodwin model, the monetary flows of the two primary loops of Fig. \ref{fig:model} are conserved. At a given time $t$ the net flow into each node is zero (Kirchhoff's Law):
\begin{align} \centering
Labor:\ & W - C = 0 \\
Firms:\ &  C + I + B - W - R - B = 0 \label{eq:firms}\\
Capital:\ & R - I = 0 \\
\Longrightarrow \quad&W = C, \quad R = I \label{eq:simple}
\end{align}
Thus, workers consume what they earn and capital owners invest what they receive. Production $Y$, or the total outflow from firms to labor and capital, is $Y = W+R$. Thus, wages $W$, or the workers' share of the production, and rent $R$, or the owners share of the production, are given by
\begin{align}
W = s_w Y &= s_w \theta_w \lambda_w N_w P \label{eq:W} \\
R = (1-s_w) Y&= (1-s_w)\theta_w \lambda_w N_w P \label{eq:R}
\end{align} 
respectively.

The rate of change of production is determined by investment, i.e. production is assumed to be proportional to the accumulated investment (Cassel-Harrod-Domar Law, combined with Say's Law, so that firms re-invest all of their profits and do not save any):
\begin{equation}
Y = \chi K = \chi \int_{-\infty}^t dt' (1-s_w(t')) Y(t') \label{eq:Y1} e^{-\phi (t-t')}
\end{equation} 
depreciated by an additional pre-specified exponential, and the rate of change of the production is
\begin{equation}
\frac{dY}{dt} =  \chi (1-s_w(t)) Y(t) - \phi Y(t)
\end{equation}  

The second dynamical variable is the workers' share of production, $s_w$. Using  
\begin{equation}
\frac{1}{Y}\frac{dY}{dt} = (\alpha + \beta + \gamma + \frac{1}{\lambda_w}\frac{d \lambda_w}{dt}) 
\end{equation} 
we can write 
\begin{equation}
\frac{1}{\lambda_w}\frac{d \lambda_w}{dt} =  \chi (1-s_w(t)) - \alpha - \beta - \gamma - \phi
\end{equation}  

\begin{sidewaysfigure}
	\centering
	\includegraphics[width=0.9\textwidth]{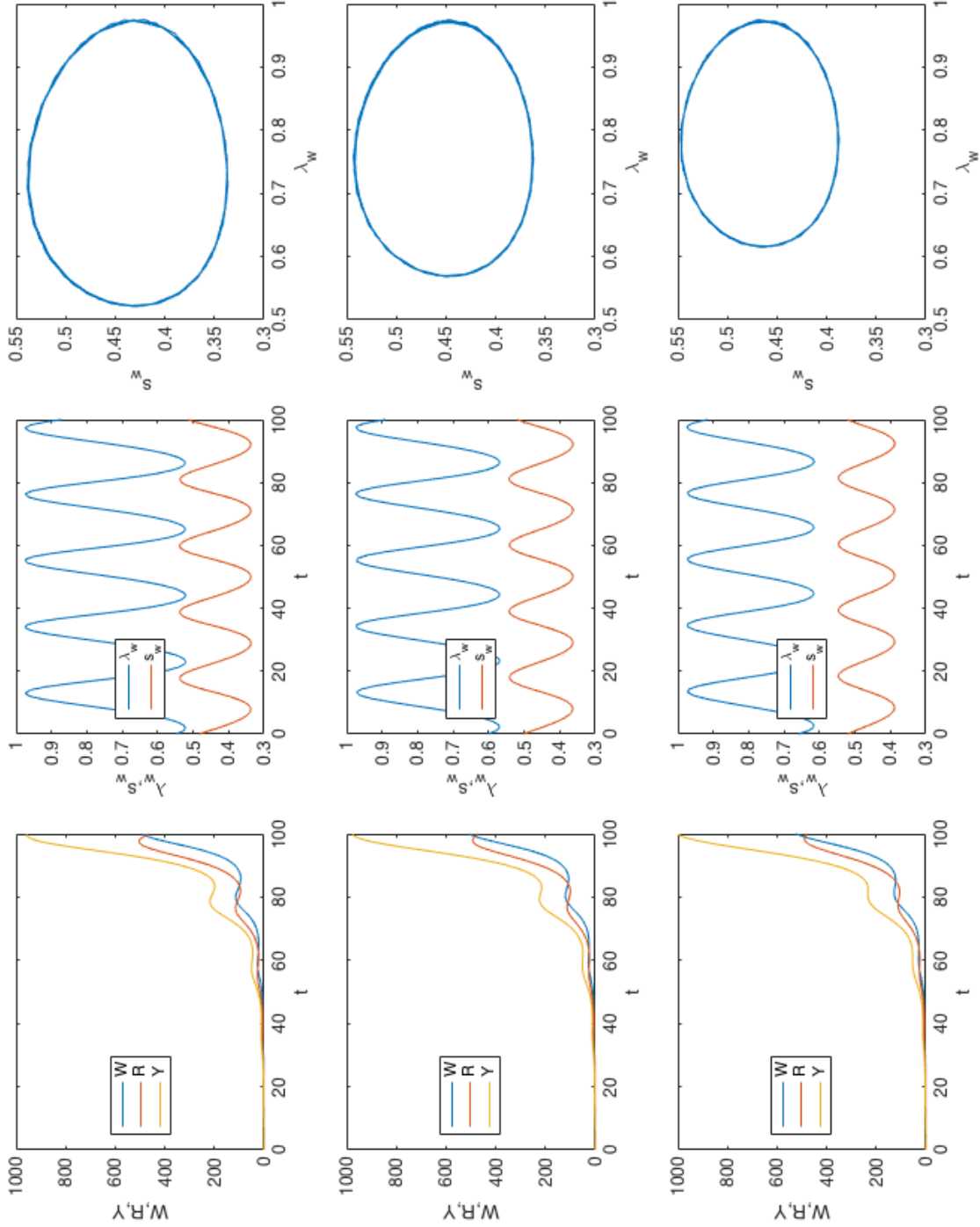}
	\caption{Representative simulations of Goodwin's model, showing wages, rent, and production (left column); employment rate and workers' share (middle column); and a state (phase) plot (right column). The initial conditions are $\lambda_w=0.85, s_w=0.75$ (top row), $\lambda_w=0.9, s_w=0.75$ (middle row), and $\lambda_w=0.95, s_w=0.75$ (bottom row). The coefficients for $\theta_w, N_w, P$ are, $\alpha = 0.01, \beta = 0.01, \gamma = 0.05$, respectively.} \label{fig:macro}
\end{sidewaysfigure}

The Goodwin model describes the dynamics of $s_w$ and $\lambda_w$. The model uses Phillips curve to relate wage inflation to unemployment. The resulting coupled differential equations, identical in form to the Lotka-Volterra equations in biology, are
\begin{align}
\frac{ds_w}{dt}& = -(a-b\lambda_w) s_w  \label{eq:lambda2}\\
\frac{d\lambda_w}{dt}& = (c - ds_w) \lambda_w \label{eq:s}
\end{align}
where $a, b, c, d$ are constants. The dynamics of these equations can be solved analytically and include oscillatory behaviors that are cycles in the space of the dynamic variables.
Representative simulations are shown in Fig. \ref{fig:macro}.  
We scaled the solutions so that the maximum value of $\lambda_w$ is 0.975 (corresponding to 2.5\% unemployment, which is the historical minimum in the US). 
While the deterministic dynamic solutions for a particular value of the parameters can be obtained, those solutions generically are highly sensitive to variation. The solutions of the Lotka-Volterra model are highly sensitive to initial conditions, and small perturbations knock the system from one periodic orbit to a very different one. The dynamical behavior of a real world system that is subject to noise, parameter changes, or modifications of mechanism that describe the dynamical behavior (i.e. the equations themselves) may only be described over limited periods of time by a solution. How long the solutions would be valid depends on the level of noise and other variations and can be characterized by a sensitivity parameter. The Lotka-Volterra equation is also not generic---the dynamical equations of a real world system should have additional terms in the differential equation, and the sensitivity of the dynamical solutions to the existence of those terms implies that they would play an important role in determining the dynamics of the system. It is not therefore a robust starting point for the dynamical description of real world systems. 

\section{Universal equations}
The assumptions that are present in the Goodwin model enable a characterization of the way economic systems relate to the mathematical treatment of the two loop model. However, these assumptions also limit our confidence that they provide the right description of the economic system as there may be a variety of effects that would change the specifics of those equations. This is particularly problematic given the sensitivity of Goodwin model solutions to variations in parameters, initial conditions, perturbations and model assumptions. We therefore consider the resulting equations rather than the assumptions leading to them and consider the properties of a two loop system that is characterized primarily by two variables: the flows in those loops.

A key insight from multiscale information analysis of complex systems is that identifying the relevant parameters is a more fundamentally important first step than the specific mechanistic or (as appropriate) stochastic dynamical equations that describe their behavior. From such an identification we can determine generic (universal) dynamical behaviors that encompass any real system, characterize distinct regimes of behavior, identify what are the essential levers for change, and use limited empirical data to identify the specific dynamics that are taking place in a particular system at a particular time. The specific parameter values of the resulting model can then be empirically derived effective parameters rather than ones that represent more specific, potentially limiting, assumptions. 

We thus frame an economic analysis primarily through a universal equation in two variables $(x,y)$, where $x$ is consumption or wages, and $y$ is investment or returns, the flows through the two primary loops. The variables may alternatively be considered logarithms of the flows: $x = \ln(C)$, $y= \ln(I)$. We restrict the analysis to a linear equation at this point, which considers the effect of changes in system variables to first order.
Why would it be helpful to describe the dynamical system only to first order? For a very large real world system the changes in system variables in a relevant time interval (here monthly or yearly) are small, here only a few percent. The small and smoothly varying changes from year to year suggest an expansion is a useful approximation. Moreover, the absence of a linear term in the expansion would be surprising in any complex system. A zero value of a parameter would happen coincidentally, at a particular time, during a shift in parameter values that changes system behavior from one regime to another, i.e. a shift from positive to negative values of that parameter. This occurs in phase transitions in materials, and can be considered as a transition in the regime of behavior of the system. It follows that the first order expansion is the most robust description of the system. Higher order terms will contribute over longer times and can be studied as refinements. 

We are limiting our assumptions to: (a) existence of two dominant nearly disconnected loops; and (b) smoothness of dynamics so that an expansion is valid to first order. Later we will add one more assumption to analyze the role of economic policy interventions: (c) economic growth.

Based upon the first two assumptions, the general form of the first order equations can be written: 
\begin{equation} \begin{array}{ll} 
d_t x &= a' + b' x + c' y \\
d_t y & = d' + e' x + f' y 
\end{array} \end{equation}
According to this equation $a'$ and $d'$ describe the rate of increase of wages and investment independent of the current value of wages and investment themselves. The factors $b'$ and $c'$ describe how much increasing $x$ and $y$ contributes to the positive rate of change of $x$, while the factors $e'$ and $f'$ describe how much an increase in $x$ and $y$ contribute to the rate of increase of $y$. It is possible that an increase in $x$ contributes to an increase in $x$, or to a decrease, depending on the sign of $b'$. Given a combination of such dependencies there can be convergence of the economic activity to a fixed point, divergence, or oscillation. The oscillation might be considered a linear version of the oscillations of the Goodwin model, though the effect of noise and other variations is different. Capturing whether there is exponential growth or oscillatory behavior and, even more critically, whether there are changes in behavior of the system over time are important aspects of the characterization of an economic system. We limit our objectives accordingly for the time being.

We can also write the same system as 
\begin{equation} \begin{array}{ll} 
d_t \tilde{x} &= b' \tilde{x} + c' \tilde{y}, \\
d_t \tilde{y} & =  e' \tilde{x} +  f' \tilde{y},
\end{array} \end{equation}
where $\tilde{x} = x - x_0$, $\tilde{y} = y - y_0$, and $x_0$, $y_0$ are stable or unstable fixed points. A discrete time equation version corresponds better to available data for 
monthly or annual measurements of economic activity:
\begin{equation} \begin{array}{ll} 
\tilde{x}(t+\delta t) - \tilde{x}(t)  &= b'' \tilde{x} + c''  \tilde{y}, \\
\tilde{y}(t+\delta t) - \tilde{y}(t)& =  e'' \tilde{x} +  f'' \tilde{y},
\end{array} \end{equation}
with $b'' \approx b' \delta t$ and similarly. This also makes apparent that the expansion is first order in the differences over time and differences relative to stable points of the dynamics, with higher order terms necessary only when these deviations become large enough. For exponential growth, the deviations from equilibrium values eventually become large, but the expansion is still a useful approximation with an adjustable exponential growth rate. The shape of large oscillations that are influenced by nonlinearities are less essential to our objectives than their existence, even if the shapes are interesting and ultimately important for a detailed analysis. Thus, the expansion is appropriate for systems with smooth dynamical behaviors on the relevant time frame, i.e. months or years---generally for systems that are large compared to perturbations.

For analysis, the six original parameters can be reduced to two by rescaling. Using the differential notation for convenience, setting $\tilde{x} = (c'/b') \check{x}$, $\tilde{y} = \check{y}$, $t = \check{t}/b' $, $e'''=e'c'/b'^2$, $f'''=f'/b'$, gives
\begin{equation} \begin{array}{ll} 
d_{\check{t}} \check{x} &= \check{x} + \check{y}, \\
d_{\check{t}} \check{y} & =  e''' \check{x} +  f''' \check{y},
\label{eq:modes}
\end{array} \end{equation}
so that the solution space is two dimensional. However, in fitting to real world behavior all six parameters may be used. The solution of the linear equations can be written directly, and outside of a minor complication that arises if there is a coincidence of degeneracy of the coefficient matrix, which would be unlikely without a symmetry in a real world system, the solution is written as:
\begin{equation} \begin{array}{ll} 
\{\tilde{x},\tilde{y}\} & = s_1 v_1 e^{\lambda_1(t-t_0)} + s_2 v_2 e^{\lambda_2(t-t_0)} \\
\label{eq:linear} \end{array} \end{equation}
where
\begin{equation} \begin{array}{ll} 
\lambda_1 &= (b' + f' - Q)/2 \\
\lambda_2 &= (b' + f' + Q)/2 \\
v_1 &= \{(b' - f' - Q)/2, e'\} \\
v_2 &= \{(b' - f' + Q)/2, e'\} \\
Q &= \sqrt{b'^2 + 4 c' e' - 2 b' f' + f'^2} \\
s_1 &= (\tilde{x}(t_0),\tilde{y}(t_0)) \cdot v_1 / ||v_1||^2 \\
s_2 &= (\tilde{x}(t_0),\tilde{y}(t_0)) \cdot v_2 / ||v_2||^2 \\
\end{array} \end{equation}

The regime of oscillation is given by the case when $Q$ is imaginary. The eigenvalues and eigenvectors, as well as $s_1$ and $s_2$, are complex conjugate pairs. This has to be true because the matrix of coefficients are real. The real part of the eigenvalues still provide an overall exponential growth or decline in the oscillatory regime. The case of pure convergence or divergence is given by $Q$ real and the larger of $\lambda_1$ and $\lambda_2$ greater or less than zero. 

It is convenient to display the dynamics by normalizing the economic flows so that the primary exponential growth is omitted. The fraction of economic activity is then given by 
\begin{equation} \begin{array}{ll} 
z &= x/(x+y)
\end{array} \end{equation}
In the exponential regime, if one exponential dominates the other, then we have
\begin{equation} \begin{array}{ll} 
z &\approx z_0 + z_1 e^{-\lambda(t-t_0)} \\
\lambda &= \lambda_2-\lambda_1 = Q
\label{eq:exp}
\end{array} \end{equation}
In the oscillatory regime, we recover an oscillation:
\begin{equation} \begin{array}{ll} 
z &\approx z_0 + z_1 \sin(k (t-t_0) + \phi_0)
\label{eq:sin}
\end{array} \end{equation}
with $k=iQ/2$.

\section{Data}

National accounting reports production in order to measure GDP in its entirety \cite{NIPA1}. We wish to separate the transactional flows of the national income and product account (NIPA) tables between labor and capital flows as shown in Figure \ref{fig:model}. Households may participate in both labor and capital activities, thus the separation of capital and labor need not be absolute. Instead, for our analysis to hold, it is sufficient that there is only limited net transfer between the two loops. For example, returns on capital assets are not substantially used for the consumption of goods, and only a small fraction of wages are used for investment. This is an appropriate hypothesis as will be shown by the approximate closure of the two-loop model. It might be suspected that the housing sector plays a special role because home owners are often participants in labor. However, it is sufficient that returns on housing investment are generally used for reinvestment in capital assets, and home purchases are generally not made outright from wages.

We categorized the transactional flows of the NIPA tables from 1960 to 2015 using the personal, firm and government income and outlays table as detailed in Table 1 and in Appendix \ref{appendix:data}.

\begin{table}[]
\centering
\caption{US government reported macroeconomic data for categories of flows shown in Fig. \ref{fig:model} \cite{NIPA1}. We modified stared items because of imputations that are not relevant for our analysis (see Appendix \ref{appendix:data} for details).}
\label{datatable}
\begin{small}
\begin{tabular}{|l|l|l|l|l|}
\hline
\bf{From} & \bf{To} & \bf{Quantity}                                     & \bf{NIPA Table} & \bf{Line} \\ \hline
 Firms      & Capital    &*Interest and miscellaneous payments        & 1.16       & 9    \\ \hline
 Firms      & Capital    & Transfer payments to persons (net)           & 1.16       & 13   \\ \hline
 Firms      & Capital    &*Proprietors' income with adjustments          & 1.16       & 16   \\ \hline
 Firms      & Capital    &*Rental income with adjustments                   &1.16       & 17   \\ \hline
Firms      & Capital    &*Net dividends                               &  1.16       & 23   \\ \hline
 Firms      & Capital    &*Undistributed corporate profits with adjustments  & 1.16       & 24   \\ \hline
 Capital    & Firms      &*Interest receipts                           & 1.16       & 4    \\ \hline
 Capital    & Firms      & Private fixed investment                      &5.3.5      & 1    \\ \hline
 Firms      & Labor      &*Compensation of employees                   & 2.1        & 2    \\ \hline
Labor      & Firms      &*Personal consumption expenditure                              & 7.12     & 6   \\ \hline
Firms      & Government &Taxes on production and imports               &  3.1        & 4    \\ \hline
Firms      & Government &Taxes on corporate Income                     &  3.1        & 5    \\ \hline
 Firms      & Government & Transfer payments from business (net)         &3.1        & 17   \\ \hline
 Firms      & Government & Interest receipts on assets         &3.1        & 10   \\ \hline
 Government & Labor      &Government social benefits to persons         & 2.1        & 17   \\ \hline
 Government & Firms      &*Consumption expenditures                    &3.1        & 21   \\ \hline
Government & Firms      & Subsidies                                     & 3.1        & 30   \\ \hline
 Government & Capital    &*Interest payments to persons and business   & 3.1        & 28   \\ \hline
Capital    & Government &*Capital personal current taxes                    & 3.1        & 3    \\ \hline
 Capital    & Government & Income receipts on assets           &3.1        & 10  \\ \hline
 Capital    & Government & Transfer receipts from persons            &3.1        & 17   \\ \hline
 Labor      & Government &*Contributions for government social insurance &3.1        & 7    \\ \hline
Labor      & Government &*Labor personal current taxes &3.1        & 3    \\ \hline
\end{tabular}
\end{small}
\end{table}

The Investment flow (Capital to Firms) includes fixed investment as well as capital borrowing costs (monetary interest payments by people and firms) which are accounted for in receipts of interests by firms, especially banks. The Return flow (Firms to Capital) includes any form of undistributed and distributed profits, including interest payments---all profit made by a firm (income minus wages and government taxes) is considered earned by the owners of the firm.

The Consumption flow (Labor to Firms) includes personal consumption expenditures (PCE) corrected to exclude non-monetary contributions to GDP such as produced but unsold goods, financial services furnished without payments, and owner-occupied housing rent. These contributions to GDP in the NIPA tables do not arise from monetary consumption flows but from reclassification of other contributions to production by the Bureau of Economic Analysis (BEA) that constructs the tables. Wages paid by foreign employers to US residents are included in the compensation of employees in the NIPA tables. For consistency, we include the consumption by US residents abroad and exclude foreign consumption in the US. We also subtract the NIPA valuation of final consumption expenditures of nonprofit institutions serving households which accounts for the value of underpriced services. These adjustments are similar to the ``market-based" PCE, which has been calculated by BEA since 1987, with a few exceptions. The amounts of all these adjustments are not significant in relation to conclusions reached. 

Flows from Labor and Capital to Government and conversely were corrected to exclude non-monetary services by and for banks and underfunded pension plan interest payments. We estimated Labor and Capital portions of taxes by using IRS Tax data for top earners. The numbers we report separate taxes at the $1\%$ level of $\$465,000$ in adjusted gross income in 2014 \cite{IRS} and take the fraction in prior years in proportion to the historical ratio of prevailing maximum tax rates, limited to $100\%$. A wide range of alternative assumptions give the same conclusions.

\begin{figure}[tb]  
	\centering	\includegraphics[width=0.85\textwidth]{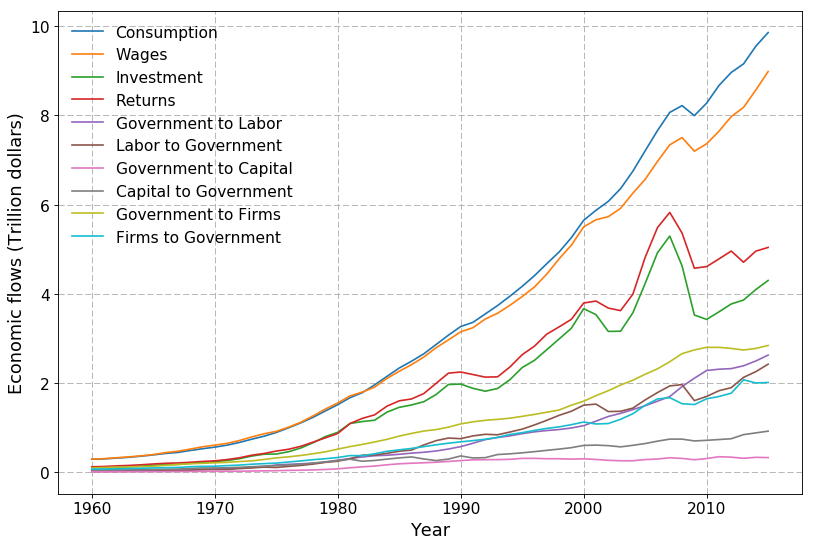}
	\caption{Economic flows in the US from 1960 to 2015 according to categories of Fig. \ref{fig:model}. Growth of GDP is reflected in all four primary flows. We can observe the association in magnitude and fluctuations of consumption with wages, as well as of investment with returns. Fluctuations are larger in investment and returns.}
	\label{fig:growth}
\end{figure}

\begin{figure}[tb]  
	\centering	\includegraphics[width=0.99\textwidth]{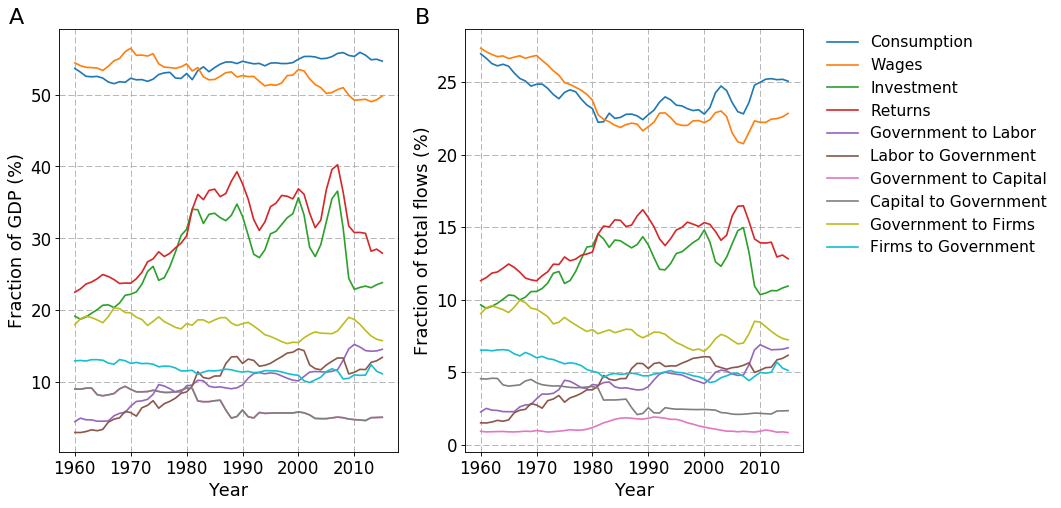}
	\caption{Fraction of economic activity for economic flows. A. Flows as fraction of GDP. B. Flows as fraction of total of flows. The dominant flows are those of the primary loops in Fig. \ref{fig:model}. As in Fig. \ref{fig:growth} we can observe the association in magnitude and fluctuations of consumption with wages, as well as of investment with returns. A change in behavior appears to occur around 1980 (see Fig. \ref{fig:fits}). }
	\label{fig:fractions}
\end{figure}

Figure \ref{fig:growth} shows the growth of economic activity in the US from 1960 to 2015 using the categories of transactional flows from Fig. \ref{fig:model}. The monetary flows normalized by GDP and by total flow are shown in Fig. \ref{fig:fractions}. Normalization eliminates the primary growth from the dynamics making more apparent the relative dynamics. We see that consumption/wages and investment/returns track each other, consistent with the hypothesis of Fig. \ref{fig:model} that these can be considered as the primary financial flow cycles. The average absolute value of deviation in the labor loop is $5\%$ and the maximum is $12\%$ when calculated in proportion to the wages. The average absolute value of deviation in the capital loop is $11\%$ and the maximum is $26\%$ (in 2010) when calculated in proportion to the returns. These deviations can be accounted for by savings, borrowing and government interactions as indicated in Fig. \ref{fig:model}. In both cases the non-closure of the primary loops is an order of magnitude less than the value of the primary flows.

\begin{figure}[tb]  
	\centering	\includegraphics[width=0.9\textwidth]{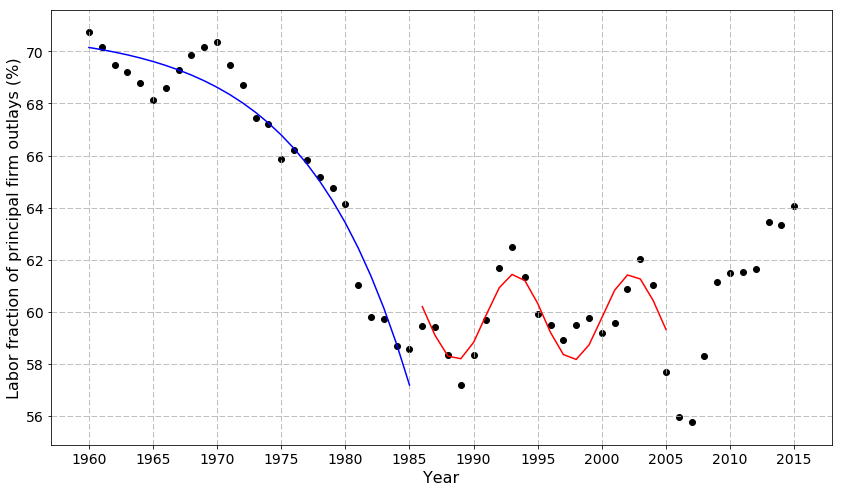}
	\caption{Wages divided by wages plus returns---percentage of economic activity in the wages and consumption loop compared to the total in the two dominant economic loops (similar to $s_w$ in Goodwin's model). Fits of exponential (blue) and sinusoidal (red) curves are shown in order to highlight the transition between different regimes of behavior that is apparent in 1980. For completeness, using the expression in Eq. \ref{eq:exp} from 1960 to 1985 we have the coefficients $\lambda=0.12 \si{/yr}$, $z_0 = 70.8 \%$, $z_1 = -0.64 \%$, $t_0 = 1960$; and Eq. \ref{eq:sin} from 1986 to 2005 we have $z_0=59.8\%$, $z_1=1.65\%$, $k=-0.69 \si{/yr}$, $\phi_0=-6.0$, and $t_0 = 1986$, with $p<10^{-15}$ and $p<0.0001$, respectively.}
	\label{fig:fits}
\end{figure}

Fig. \ref{fig:fractions}  suggests the presence of distinct regimes, and particularly a transition between regimes in 1980. Including recent times, the behavior can be described to first order by three regimes: an exponential regime until 1980, an oscillatory regime until the financial crisis, and a third regime since then for which a short time has passed and the ability to characterize is limited. The transition in 1980 is readily apparent in Fig \ref{fig:fits} which shows the proportion of the labor loop to the overall value of the principal loop flows as measured by wages over wages and returns. The behaviors in both regimes are consistent with the first order normalized Eq. \ref{eq:linear} and simplified versions Eq. \ref{eq:exp} and Eq. \ref{eq:sin} as shown by the fitted curves. We note, however, in advance of further analysis described below, that the dynamics described by these equations is not endogenous to the economy as it is linked to regulatory changes in interest rates.

The first regime from 1960 to 1980 has an increasing importance of the capital loop compared to the labor loop. The second regime, from 1980 to 2007, is marked by oscillatory competitive interactions between the labor and capital loops. 

In seeking a reason for the existence of two regimes, we should consider how they may be related to a shift in US Federal Reserve bank monetary policy \cite{1980policy}, as well as changes in fiscal policy including tax rate changes \cite{reagan} in the early 1980s. Understanding the impact of regulation and policy changes requires an understanding of the mechanism of changes in the primary economic loop flows, particularly how they grow.

\section{Economic growth and monetary flows}

Economic activity is naturally dynamic. Population growth, improvements in productivity, changes in technology, accumulation of assets and other changes affect the rate of economic activity over time. For this reason, we expect the flows to change over time. Investment is needed in order to overcome depreciation as well as respond to other sources of change, such as growing demand due to population increases and technology based improvements.

While it is possible to imagine and even identify historical cases where economies shrink,
we will focus on a growing economy. The overall GDP having increased by an order of magnitude in 40 years (Fig. \ref{fig:growth}); this is a reasonable starting point for the US economy. A growing economy implies that the circulation in loops increases. The circulating flow is determined both by the amount of money and rate at which transactions occur around the loop. The rate of flow is a product of the two. It might be imagined that synchronous transfer of money around the loop can increase flow without increasing the amount of money, however, this scenario does not apply due to built-in delays in transaction processing. Since we are considering multiple cycles of the economy, measures of overall money supply (direct or indirect) and the aggregate money velocity measures are not likely to be of help in our analysis \cite{money,velocity}. For example, savings are included in traditional measures of M3 and influence the velocity of the ratio between nominal GDP and money supply.  However, in our analysis, savings themselves are not involved in flows, and changes in savings in and of themselves do not directly affect the velocity. Still, we recognize that such monetary measures are a step in the direction of evaluating the flows that are of importance in our model.

Growth of the economic activity implies that either the amount of money circulating in a loop or the velocity must grow accordingly. It is reasonable to expect, however, that, barring dramatic changes in the nature of economic transactions, the rate of labor payments at weekly, biweekly or monthly installments sets the relevant time scale of wages and consumption loop transactions and can change somewhat but not by an order of magnitude. How the rate of investment transaction has changed is less transparent, however only the net investment in production equipment and property is relevant to this loop, not buying and selling per se (e.g. of securities), the rate of which has increased dramatically. Thus, we concentrate on the process of increase of the amount of money, which may change dramatically over extended time and is generally understood to be the regulator of economic activity. 

For an extended growth period, growth of the money supply must generally be by the same factor in both of the two primary economic cycles, with variations in their relative growth rate of limited duration. Otherwise only one of the cycles will be present. This would mean that either there is no wages and consumption, or no investment and returns. The former would mean no economic activity, which is inconsistent with having investment and returns. The latter would mean no ability to overcome depreciation or accommodate technological changes, which also seems impossible.

\section{Monetary Injection and Borrowing}

An increase of currency circulating in the labor and capital loops can originate from either the combined effect of Government tax policy, entitlements and spending, or through labor and capital borrowing. From a regulatory standpoint, the former is the effect of fiscal policy, and the latter is linked to monetary policy setting the interest rates of borrowing. We call the former fiscal injection and the latter borrowing.

When fiscal injection occurs preferentially into one of the two loops, it is natural for that loop to save the excess. In order for a balanced growth to take place in the two loops, the other loop has to make up for its deficit, by borrowing and accumulating debt or by direct transfer from the other loop from corporate wage and return policy. If fiscal injection is predominantly into the Labor loop, Labor can both increase consumption and save. Moreover, it is natural that Capital will borrow to take advantage of opportunities to sell more to Labor. If excess money is injected into the Capital loop, Capital can invest and save. Under conditions when Labor does not have enough money to consume its needs or desires, it will borrow. Borrowing may be linked to interest rates set by monetary policy so that the balance between the loops may not be achieved endogenously. 

\begin{figure}[tb]  
	\centering	\includegraphics[width=0.99\textwidth]{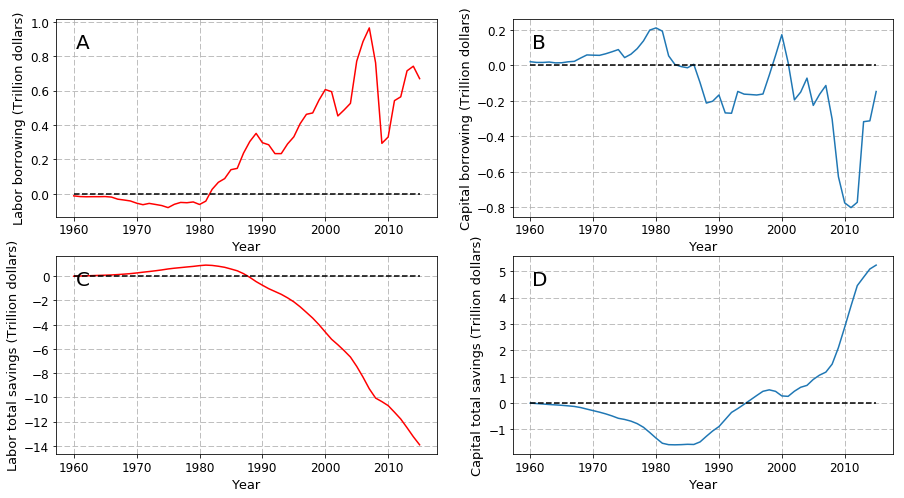}
	\caption{Estimates of borrowing and total savings (or debt) for Labor and Capital. A transition from capital borrowing to labor borrowing and capital savings in 1980 is evident. A.  Labor borrowing obtained by subtracting wages and government benefits from consumption and taxes. B. Capital borrowing obtained by subtracting returns and government interest payments from investment and taxes. C. Labor total savings obtained by aggregating borrowing since 1960 and D. Capital total savings obtained by aggregating since 1960. Total savings (debt) is obtained by aggregating borrowing since 1960.}
	\label{fig:debt}
\end{figure}

Therefore, we can identify which regime the system is in by looking at the savings and debt associated with the two loops. Savings are associated to excess from fiscal injection, while the other loop accumulates debt. Measures of borrowing by Labor and Capital are shown in Fig. \ref{fig:debt}. Both curves are consistent with the existence of a change in regime in 1980. Labor (consumers) were able to save before 1980, while after 1980 there is a progressive increase in the rate of borrowing, which increased dramatically by the financial crisis (Fig. \ref{fig:debt} A and C). In contrast, capital owners borrowed increasing amounts before 1980, and savings increased afterwards (Fig. \ref{fig:debt} B and D). Consistent with the labor borrowing, wages are higher than consumption before 1980 and lower afterwards. Returns are higher than investment except for a brief period around 1980, indicating that prior to 1980 the higher returns were not sufficient to satisfy the investment needs of capital owners. In summary, this suggests a shift from fiscal policy injecting money primarily into the labor wages and consumption loop, to being injected into the capital investment and returns loop. 

The conclusion is that until 1980 we have government flows predominantly to the Labor loop and subsequently to the Capitol loop. Borrowing occurred first by Capital and then by Labor during these periods. The existence of two regimes prior to and after 1980 is consistent with fiscal policy changes in the early 1980s, including a reduction of taxes for high income persons \cite{reagan}. The changes are readily understood to shift the flow of Government monetary injection from Labor to Capital. A connection to shifts in methods of adjusting interest rates in response to inflation by the US Federal Reserve bank policy at that time \cite{1980policy} seems less apparent.

The compensation of the fiscal injection to Labor and Capital by borrowing in the other loop is consistent with a system adjusting itself to compensate for an imbalance. This is a general principle applicable to systems that are in equilibrium or have self-consistent dynamical processes. When we shift a quantity away from an intrinsic balance by external force, the system responds to bring it back toward balance. 

Examples abound in physical and chemical science including the screening of electric field in electro-magnetics by a dielectric, the stress-strain response, and Le Chatelier's Principle in chemistry. The dielectric response of a system reduces but does not eliminate the external electric field $E$ resulting in the net electric displacement field ($D=\epsilon E$); the strain of a material in response to an external stress acts to reduce the stress (traditional assumptions often assume a specified stress but that is a specific assumption) but does not eliminate it. 

This is consistent with the treatment of the economy as a system that deviates from its optimal value and adjusts through internal dynamics to bring it closer to that optimal value. The extent to which it succeeds to approximate the optimal value depends on the response mechanism. In economic regulation this effect is sometimes called `headwinds' [10]. 

Typically, the net change of the system after system response is reduced compared to the impact of the original shift. The residual of the external force is a net force that self-consistently causes the system to adjust. In economics the behavior may be distinct from these other contexts due to its intrinsic non-equilibrium behavior. For a non-equilibrium system it is possible for a higher than unit response, due to internal amplifying or cascading dynamics. We will see that this is important in this context. However, we will also find that during this period the endogenous response of the system does not describe the observations and that regulatory action is involved.

\section{Regulation and control theory}

Policies associated with the regulation of economic activity can be framed similarly to diverse other regulatory systems. In a control theory discussion, the deviation from a desired state, measured automatically or by human observers, is corrected for by an externally applied force that shifts the system toward that desired state.

In standard economic regulation, federal reserve interest rates are adjusted to regulate measures of inflation. The interest rate is adjusted to be proportional to the difference of the desired and current inflation, with the proportionality factor based upon historical response according to the Taylor rule \cite{taylor1993discretion,taylor2}.
The regulatory effect is understood to occur through changes in borrowing and thus injection of money into the economy. 
Thus, the impact of this money injection on labor and capital will appear in their savings and borrowing which exposes where monetary flows are directed. Other objectives, including particularly unemployment rate, may be considered in the regulatory adjustments.

While the existence of borrowing as a means of monetary policy is more widely recognized, the injection of money by government spending also contributes to these financial flows and unlike borrowing does not bear interest and return payments. The difference between monetary injection without interest and borrowing can be important for the regulation of economic activity as the cost of borrowing contributes to flows over time.

What is generally assumed in analyses of monetary policy is that there is a single adjustable regulatory factor, the federal bank interest rate. This is true whether simpler linear response models or elaborate economic models with hundreds of equations are used to determine how to adjust that factor \cite{Solow01021956,taylor1993discretion,taylor1995monetary,brayton1996guide,english2015federal,brayton2014optimal}. It is understood that regulating both inflation and unemployment may work well if both are optimized at the same level of economic activity, or there may be a tradeoff due to a compromise in optimization between them. This, however, is not the only possible effect of multiple variable optimization.

In our analysis of economic activity, we are considering a two-dimensional system, the flows through the labor loop and the investment loop, while the regulatory function of monetary policy only has a single variable, the interest rate. Control systems can work in more than one dimensions of a system if the observation of the system recognizes the direction of change from a desired system state in each dimension and the controller has the same number of dimensions of influence. This means that the number of relevant parameters of this system's `universality class' is equal to the number of control parameters \cite{bar2016big}.

What happens when there are actually two system variables (as indicated by the Goodwin model) and only one control parameter? Ashby's Law of Requisite Variety applied to control systems states that the number of dimensions of an effective controller must equal the number of relevant dimensions of the regulated system \cite{ashby1956introduction,ashby1958requisite}. This principle was stated earlier by Jan Tinbergen in discussion of the regulation of economic systems \cite{tinbergen1952theory}. The natural outcome of a one dimensional controller regulating a two dimensional system is that only one of the variables is controlled and the other deviates from the desired control point until it reaches a natural limiting value far from the desired condition. The results are consistent with the historical instability of economic systems over periods of 20-50 years and the difficulty of effective regulation despite efforts to do so \cite{reinhart2009time}. Indeed, if economies could be regulated by a single dimension of regulation, the historical instability would be surprising; the existence of multiple regulatory dimensions provides a natural explanation. 

Our analysis suggests that using the single economic regulatory factor drastically limits the effectiveness of that regulation. We discuss this in greater detail in the next section and confirm empirically in the following one that the two dominant flows in Fig. \ref{fig:model} are relevant dimensions for macroeconomic activity, including recessions. 

\section{Insufficiency of economic regulation}

If there were only one flow then economic regulation would be consistent with the standard model. If there were currents that fed each other, i.e. a figure eight that topologically is a single loop, that would also be the same. But when there are two variables that depend on each other for growth then we have to consider these variables and particularly the difference (ratio) variable. 

The dynamic variables $x = \log(C)$ and $y=\log(I)$ can be considered as the system's degrees of freedom. It is, however, natural to switch to $u = (x+y)/2 = \log(\sqrt{CI})$, and the deviation $v = x-y = \log(C/I)$, and write
\begin{align}
d_t u &= d_tC/C + d_tI/I = \eta \\
d_t v &= d_tC/C - d_tI/I = \zeta
\end{align}
Regulating only $u$, or equivalently $\eta$, would be consistent with a conventional framework of regulating economic growth to achieve the right level of inflation. However, according to our analysis any long-term deviation of the second variable,  $\zeta$, from zero leads to divergence of the state of the system from consistent growth.  

While optimal economic activity requires a certain ratio of monetary flows between the two loops, monetary policies focused only on the amount of money flowing into the economy would not regulate the relative input into each of the two flows. The effect of a policy can therefore be characterized by the ratio of the flows that is observed in the economy. Absent a change in policy, a bias of flows towards either labor loop or capital loop would persist over time. The derivative of the ratio, $\zeta$, would therefore tend to be significantly greater or less than zero depending on which loop is growing faster. We note that the flows directly associated with monetary and fiscal policies may not be the total flows because of the internal dynamics and response of the system.

More precisely, as economic activity increases a specific trajectory, $\zeta(C)$, representing a balance between $C$ and $I$, would optimize economic growth, and higher or lower values represent two distinct regimes of suboptimal growth separated by the trajectory of optimal growth. Over a period of time, a constant slope may be a good approximation to the optimal growth trajectory.

Thus, starting from an effective economy, there are two directions of trajectories in the overall state (phase) diagram of the system. One in which the logarithmic growth of the consumption to investment flow ratio, $\zeta$, is too negative and one in which it is too positive. Either way, the economy becomes ineffective over time. Unless the rate of growth of the flows is calibrated, there will be a progressive growth of the labor loop relative to the capital loop or vice versa which will reduce economic growth.

\begin{figure}[!tb]  
	\centering	{\includegraphics[width=0.98\textwidth]{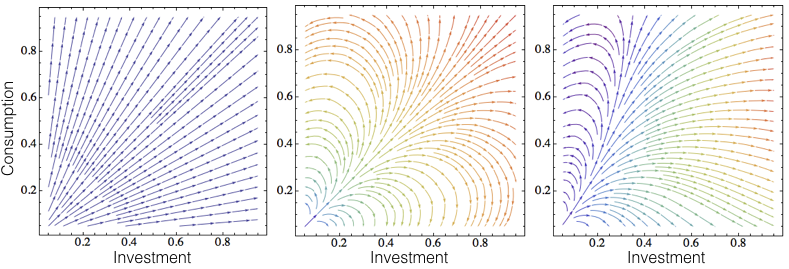}}
	\caption{Diagrams of economic development trajectories in terms of the wages/consumption (vertical axis) and investment/returns (horizontal axis). A. Shows consistent growth trajectories at a fixed ratio of consumption to investment when policies infuse the same proportion of money into each of the loops. The range of possible ratios is not limited in this diagram, i.e. it assumes all ratios can be effective. B. Shows economic growth trajectories where one of the loops becomes larger than the other, economic activity is unbalanced and declines consistent with the expectation that each of the loops is necessary for the other. C. Trajectories that arise when policies (of B) are changed to increase monetary injections to the investor loop. Scales are arbitrary.}
	\label{fig:flowstate0}
\end{figure}

Another way to analyze this ratio is to plot the consumption as a function of investment, a state space representation. A stable ratio over time means that the economy follows a straight line from the origin to the current state of the economy. According to our discussion, as the deviation from that straight-line increases, the economy becomes ineffective. The loss of economic effectiveness may be gradual or it may undergo singularities where the growth vanishes and economic activity declines at the point where one of the loops cannot sustain the flow in the other. Regulation of the economic activity by the control variable $u$ stalls when the deviation from balance in $v$ results in conditions that cannot be overcome by regulation of $u$. A change in economic policy would add an additional vector field to the flow behavior. Singularities may arise from effects such as accelerating inflation (for  $\zeta>0$), accelerating unemployment or social instability (for  $\zeta< 0$). While the presence of singularities is appealing phenomenologically, neither the presence or properties of such singularities are essential for the underlying recognition that long-term deviation from $\zeta = 0$ leads to suboptimal economic growth. Fig. \ref{fig:flowstate0} shows multiple trajectories of the economy in that phase space depending on the stability of a ratio or of regulation policy. 

Given a certain fiscal and monetary policy, the behavior of the ratio will follow one trajectory of these diagrams: Only policy changes can significantly modify the trajectory of the ratio of the economic flows. 

The effect of policies can be more complex if the policies are adapted to economic conditions over time, as they are with the Federal reserve monetary policy.

\section{Data on economic regulation}

\begin{figure}[!tb]  
	\centering	{\includegraphics[width=0.60\textwidth]{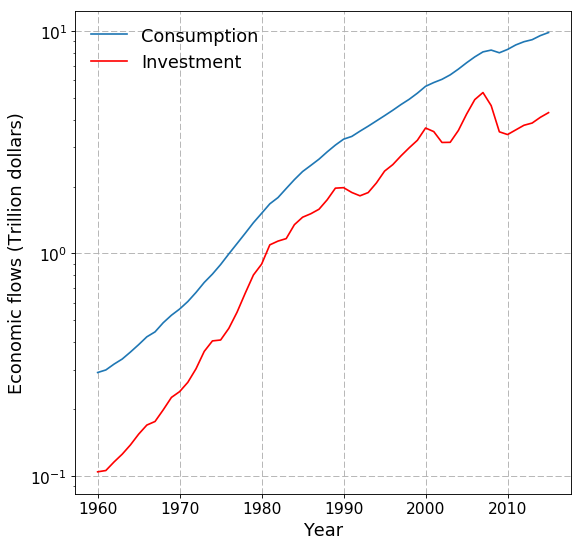}}
	\caption{Consumption and investment on logarithmic axis corresponding to dynamic variables $x$ and $y$. Consumption is generally smoother than investment.  
	}
	\label{fig:xy}
\end{figure}

\begin{figure}[!tb]  
	\centering	{\includegraphics[width=0.99\textwidth]{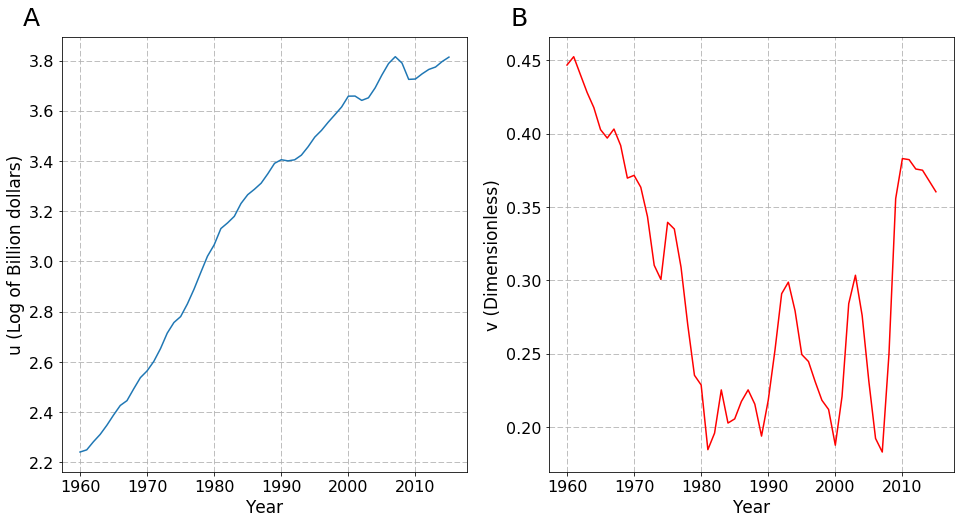}}
	\caption{Plot of economic flow variables A. $u=(1/2) \log(CI)$ and B. $v=\log(C/I)$. The latter shows the transition in 1980 and variations that are approximately periodic since then.}
	\label{fig:uv}
\end{figure}

The values of $x$ and $y$ are shown in Fig. \ref{fig:xy}, and the values of $u$ and $v$ are shown in Fig. \ref{fig:uv}. It is apparent that a shift in 1980 occurs from a negative value of $\zeta$ to an oscillating $\zeta$ since then. This implies that the growth of the investment loop was greater before 1980 and the consumption loop and investment loop growth were roughly equal afterwards. However, we will see that the oscillations incorporate the effect of policy interventions, and are not due to an intrinsic dynamic. 

\begin{figure}[!ht]  
	\centering	{\includegraphics[width=0.95\textwidth]{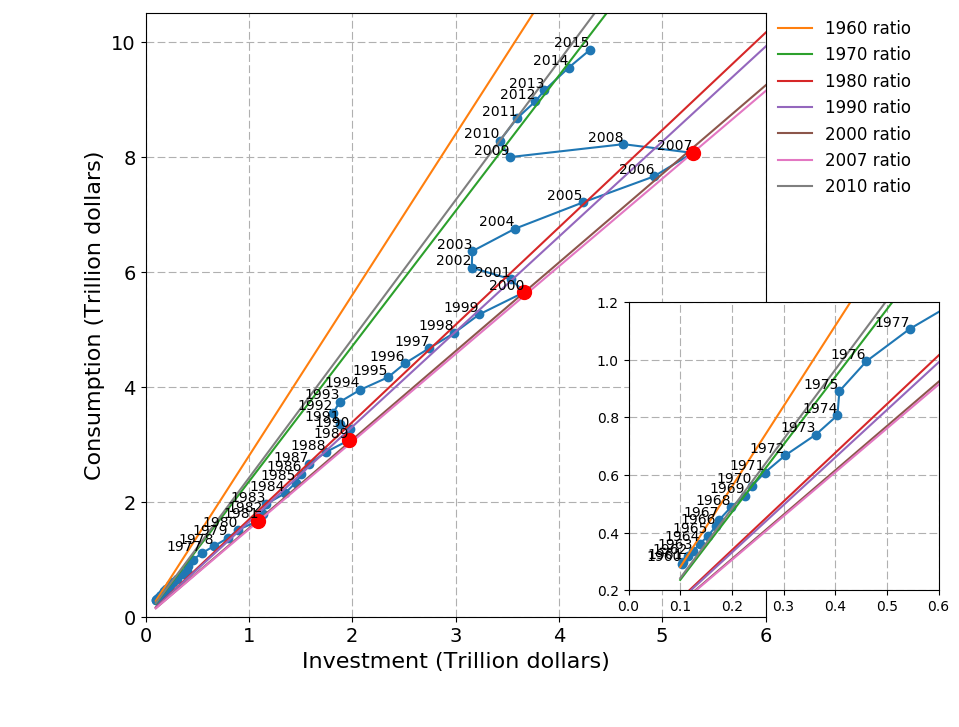}}
	\caption{Plot of consumption versus investment between 1960 and 2015. The straight lines represent the dynamics of the economy if the ratio of consumption to investment were fixed. Recessions occurred in years marked by red dots. The ratio of consumption to investment at the time of recessions is consistently $C/I=1.5$ indicating the need for a higher ratio for effective economic activity. When investment increases to $2/3$ of consumption recessions are triggered and monetary policy interventions are used to restore economic growth. The consistency of the ratio further indicates that the two flows are the dominant relevant parameters of macroeconomic activity. Otherwise variation in other economic parameters would influence this critical ratio.}
	\label{fig:CvsI}
\end{figure}

The US Economy trajectory in the state space is shown in Fig. \ref{fig:CvsI} and has an oscillatory behavior after 1980. In these coordinates, maintaining the ratio of consumption to investment is represented by following a straight line from the origin at a fixed angle to the axes. In general, the capital loop growth outpaces the labor loop growth except for certain periods of time which all follow a recession year in the early 1980s, the early 1990s, the early 2000s, and the financial crisis in 2007. 

\begin{figure}[!ht]  
	\centering	{\includegraphics[width=0.99\textwidth]{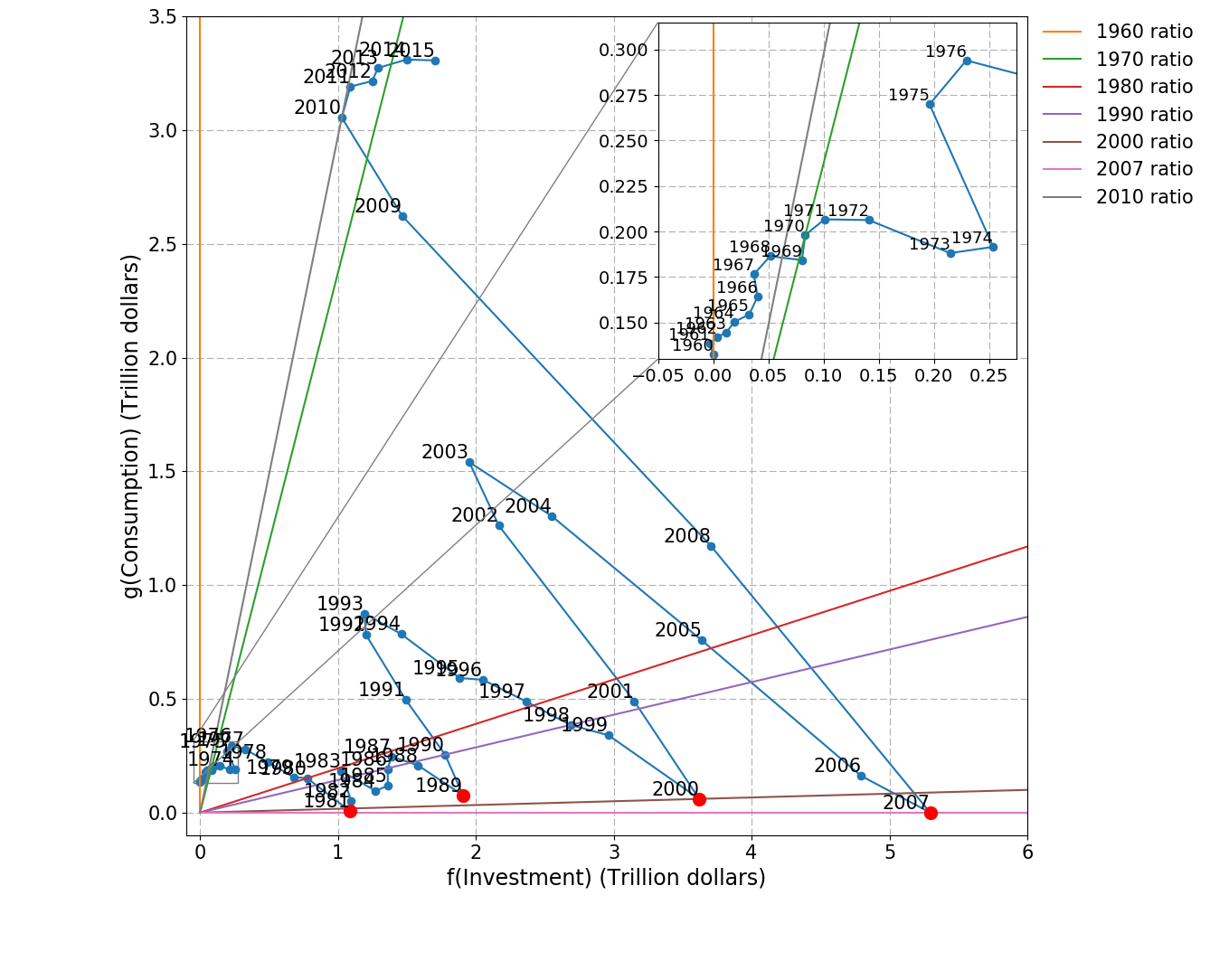}}
	\caption{Same data shown in Fig. \ref{fig:CvsI} but the region of the data between the 1960 and 2007 lines is expanded to the entire first quadrant by setting the 2007 vector direction as the x-axis (by subtracting it from all data) and, similarly, the 1960 straight line as the y-axis. Recessions occurred in years marked by red dots.}
	\label{fig:CvsI2}
\end{figure}

These properties can be better seen in Fig. \ref{fig:CvsI2}. In this figure, we transform the space in a way that preserves straight lines, but expand to a right angle the cone formed by the 1960 and 2007 lines.

The economic activity appears to bounce if it reaches a certain ratio of $C/I = 1.5$. The points at which $v$ achieves minima in the coordinate adjusted space of Fig. \ref{fig:CvsI2} occur during years of recession. Increases in the ratio of consumption to investment (by decreasing investment) are correlated with recessions and the regulatory action in response to those recessions. Absent these events, the economy has a negative $\zeta$, and investing grows more rapidly than consumption. 

The observation of recessions at a certain ratio of $C/I$ confirms multiple aspects of our analytic assumptions and inferences. It confirms that the two loop economic model, motivated by the Goodwin model, characterizes relevant parameters for economic activity. At finer levels of resolution in describing the economy other flows and variables will become relevant \cite{bar2016big}, but in describing economic recessions the results confirm there are at least two relevant parameters. It also confirms the Tinbergen/Ashby Law analysis of the importance of those parameters for regulatory action. 

The figure suggests that the recent history of economic recessions  
(early 1980s recession, early 1990s recession,  early 2000s recession, 2007 financial crisis)  is linked to an underlying dynamical behavior of shifting proportions of labor and capital flows. However, inconsistent with the framing of the Goodwin model, these oscillations do not appear to be endogenous. 

\begin{figure}[!tb]  
	\centering	{\includegraphics[width=0.95\textwidth]{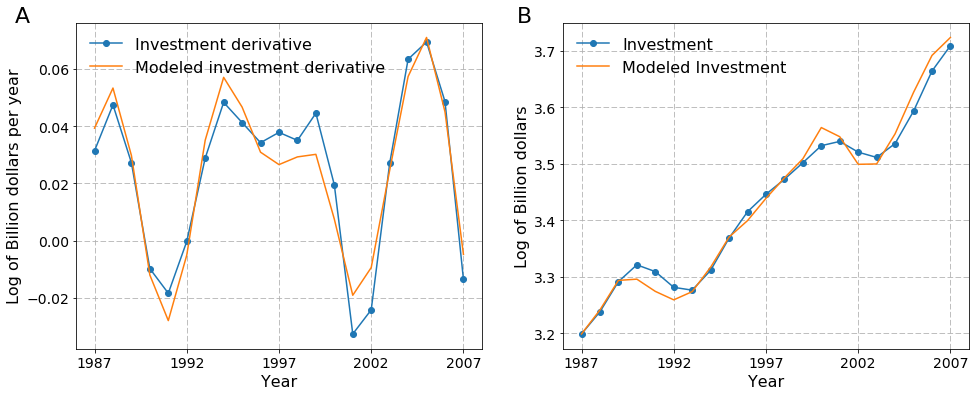}}
	\centering	{\includegraphics[width=0.95\textwidth]{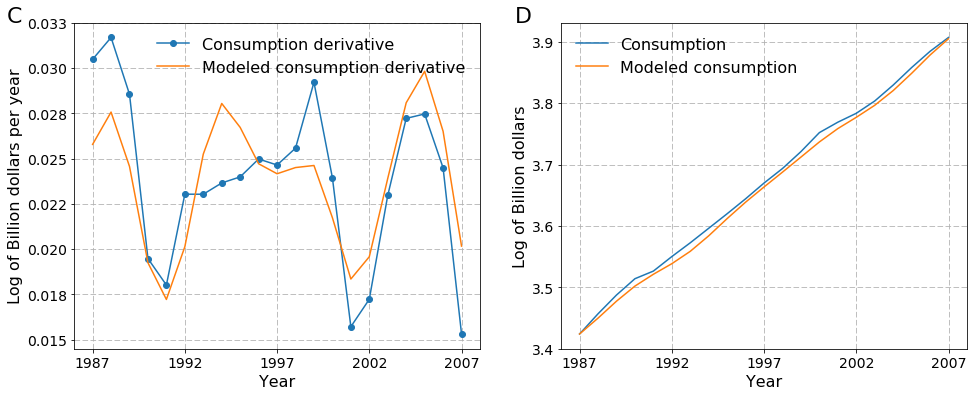}}
	\caption{Investment and consumption compared to a linear model linking them to the changes in interest rates. A. Change in log investment compared to model. B. Investment compared to model. C. Change in log consumption compared to model. D. Consumption compared to model.}
	\label{fig:ifits}
\end{figure}

Instead, the behavior can be linked to monetary regulatory action in response to the recessions by reducing interest rates. We can see this in Fig. \ref{fig:ifits} which shows the results of a model that equates changes in consumption and investment to interest rate changes. The fits are determined from a linear equation relating the change of the log of consumption to the change of interest rates $d_t \log (C(t)) = a + b d_t i(t)$ and similarly for $I(t)$.  Decreases in interest rates are related to decreases in investment and decreases in the rate of consumption growth. While this appears to be a causal model, the direction of causation should be carefully evaluated and conclusions are limited by the oscillatory nature of the overall behavior, as oscillations may be causally linked in multiple ways. Still, a connection between interest rate interventions and economic changes after recessions is justified by the conventional economic understanding of those interventions.

The observed dependencies are consistent with the interpretation that regulatory action by reducing interest rates boosts economic activity to restore growth during a recession. Thus the decreases in investment and in rates of consumption growth during a recession are alleviated and ultimately reversed by the intervention. The boost provided is temporary as the repetition of the pattern of oscillation shows. Interpreted overall, the behavior suggests that the economic activity with static fiscal and monetary policies is not consistent with a long-term zero $\zeta$, but rather an underlying negative $\zeta$ and a series of regulatory interventions (reducing interest rates) that restores growth temporarily. During the combined effects of the recession and intervention the direction of the economic trajectory changes by over $90^\circ$ (Fig. \ref{fig:CvsI}). The average movement of both consumption and investment that preserves the ratio between them is related to these periodic shifts of the direction in state space, reminiscent of sailboat tacking (Figs. \ref{fig:CvsI} and \ref{fig:CvsI2}). 

\begin{figure}[!tb]  
	\centering	\includegraphics[width=0.8\textwidth]{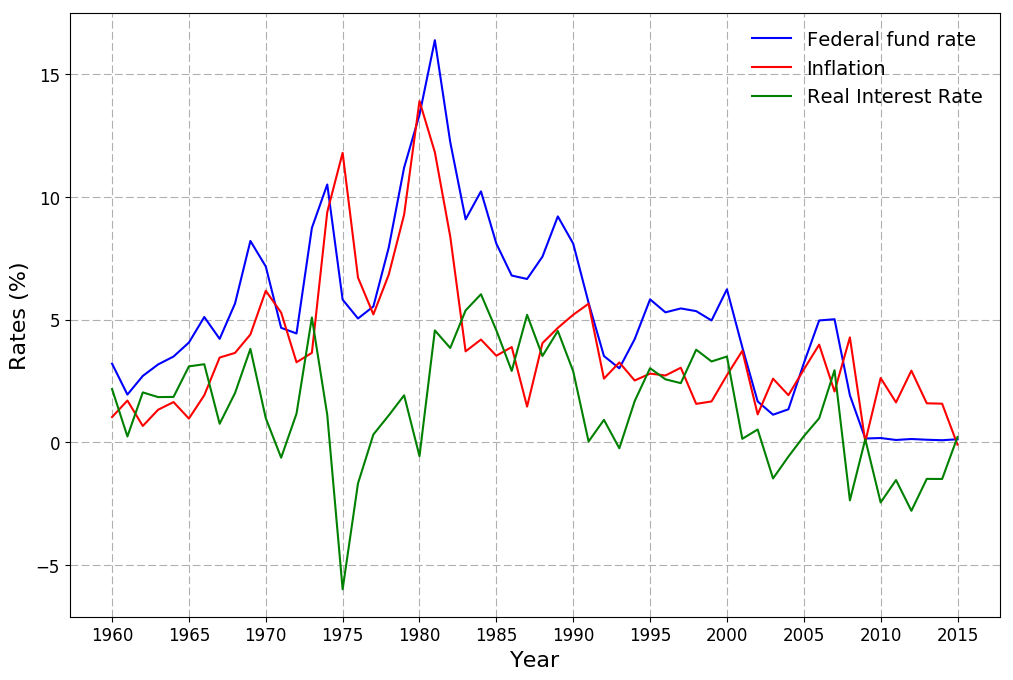}
	\caption{Interest rate (blue), inflation rate (red), and real interest rate (green) showing the two regimes of behavior prior to and after 1980 consistent with investment limited and consumption limited regimes.This suggests that the current zero interest rate is not due to the financial crisis but rather due to the limiting behavior associated with the consumption limited regime that started in 1980.}
	\label{fig:interest}
\end{figure}

The net effect of the regulatory action that stabilizes the economy is an overall trend downward in interest rates, i.e. consistent with a loss of about $12\%$ in interest rates over the period from about 1982 till 2012, $4\%$ per decade (Fig. \ref{fig:interest}). This is the average rate of interest rate reduction necessary to maintain the economic growth in a direction of constant $v$, i.e. zero $\zeta$.

There is no evidence from the observed dynamics that the economy would restore itself from recessions. Indeed, the trajectory termination on a single vector at which point monetary interventions are performed suggests that interventions become viewed as necessary, consistent with the underlying picture of departure from economic growth along trajectories that sustain ratios of flows through the two loops. Further deviation that would give rise to larger amounts in the investment loop are inconsistent with economic growth. 

We note that regulatory action is considered temporary and is designed to increase monetary flows by injection of currency into the economy by reducing the cost of borrowing. However, since monetary flows increase as part of economic growth, borrowing must also increase consistent with economic growth in order to account for the injection of currency into economic flows. Thus the total of borrowing must consistently increase with the size of the economy, including making up for any returns and interest payments. Economic activity must therefore be either subsequent to borrowing by Labor and Capital, or by Government that can inject money into the economy without direct interest or returns.

The data also suggests a different picture than is generally accepted about regulation by the Federal reserve. The usual Taylor rule picture, assumes there is a ``long run" (equilibrium) interest rate for consistent growth. Deviations from that growth can be corrected by appropriate adjustment of interest rates.  Instead, the data indicate that growth since 1980 requires ongoing reduction in interest rates. The rate of change of interest rates required $di_0/dt$ can be estimated from the observed values to be $4\%$ per decade. In this view, the reduction of interest rates in response to recessions restores growth. However, the subsequent increase of that interest rate is causally linked to the next recession. The Taylor rule assumes a one dimensional first order linear response dynamics of growth. The observed dynamics is higher order, consistent with the existence of two parameters of control. Controlling only a single parameter leads to a progressive shift of that parameter until it exceeds limits determined by properties of the system. 

Indeed, we see that the interpretation of current zero interest rates as a response to the financial crisis is misleading. It appears to be instead a consequence of the overall trend of lower interest rates since 1980. In the aftermath of the financial crisis, the federal reserve adopted new ways to inject money into the economy by direct investment termed ``quantitative easing" \cite{QE} due to the inadequacy of zero interest rates. From our analysis, the expectation that interest rates can be increased above zero \cite{FOMC} and preserve economic growth will be temporary. 

Finally, we see that in 1981, the recession occurred at the same ratio of consumption and investment as the later recessions. Thus, the economy ran into a functional boundary in 1981 associated with excess investment. This occurred despite the predominance of flows into the Labor loop prior to the recession and resulted from Capital borrowing rather than fiscal monetary injection. 
Thus, the first collision with the boundary is different in that the investment was derived from borrowing whereas afterwards investor savings accumulated. 
The recession and subsequent reduction of interest rates resulted in separation of the economic state from the boundary. The monetary stimulus is linked to increasing consumption at a time of decreasing investment, and then to increasing investment. Subsequent economic dynamics consisted of increasing investment till the boundary was reached again, and federal reserve interventions to promote consumption and later investment. 
The monetary interventions that were used to promote economic activity did not address the underlying dynamical imbalance, but provided a time of intermediate growth.  
Each interest rate reduction results in a stimulus that enables the economy to grow until it exhausts the stimulus effect by running into the boundary. 

A central conclusion is that there is a certain proportion of the wages that can be effectively invested. Over the past 30 years this proportion is about $2/3$. When investment exceeds that proportion, the economy goes into a recession. To avoid recession more money is needed in the labor loop. More generally, excess of injection into the consumer loop leads to inflation, and excess of injection into the investment loop leads to recession. 

\section{Policy and system behavior with deviation}

Our analysis of the macroeconomic dynamics has pointed to the relevance of deviations from parity in the monetary injection into the labor and capital flow loops. Deviation from parity leads to underperforming economic activity, a system response to the imbalance, and a need for periodic regulatory interventions whose utility will decrease or become ineffective. 

\begin{figure}[!tb]
	\centering
	\includegraphics[width=5in, angle=270]{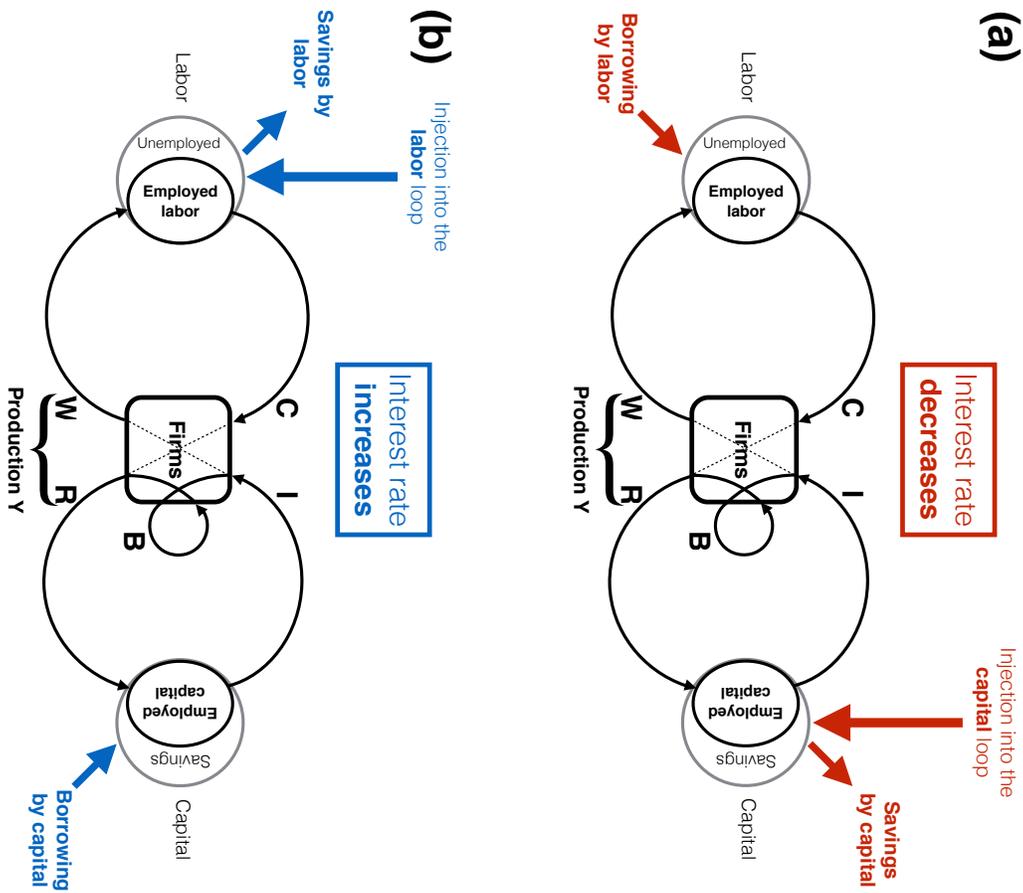}
	\caption{A schematic of responses, showing injections into (a) the capital loop and (b) the labor loop. \label{fig:injections}} 
\end{figure}

Fig. \ref{fig:injections} summarizes this schematically. According to this mechanism, injecting money preferentially to one of the wages/consumption or investment/returns loops, results in a system response that should reduce that discrepancy. For example, the response consists of borrowing on the part of those who do not have enough, and savings on the part of the ones who have too much. We can thus identify two different domains of behavior of the system: investment limited and consumption limited. 

{\it Investment limited}---Money injected into the labor loop promotes consumption, savings by labor and borrowing by capital. The borrowing is in order to take advantage of the investment opportunities that arise from the increasing and unmet demand from labor for additional consumption. However, limited injection into the investment loop implies that production is not sufficient to meet demand. This means that consumer prices increase (inflation) driven by the increase in money in the hands of consumers relative to the availability of products. The impact on interest rates must be analyzed according to its use as a control mechanism. If, as is conventional, interest rates are used primarily to control inflation in consumer products, interest rates will go up. The increase of interest rates not only draws money out of the consumption loop into savings, but also makes investment less favorable, drawing money out of investments into savings. Nevertheless, empirically, the borrowing for investment in this regime appears to be sufficient for investment to grow more rapidly than is necessary. This is consistent with a limited utility of the interest rate as a control mechanism. 

{\it Consumption limited}---When money is injected into the capital loop, there are increases in investment but the extent to which this occurs is limited by the availability of investment options that provide return greater than the interest rate. Because the demand is limited due to the lack of money in the labor loop, investing is not motivated because of the absence of expected gains. There is savings by capital and borrowing by labor (accumulating consumer debt) in order to obtain what labor considers essentials or to respond to new products. Interest is progressively adjusted lower to promote consumer spending and investment, but with low injection into wages relative to capital, response is weak due to limited resources for consumption. Instead, the reduction of interest promotes borrowing for immediate consumption. The resulting consumption motivates investment, which may also exploit progressively decreasing opportunities for return. Under these conditions inflation is more weakly coupled to monetary injection as the additional money compensates for deficit of consumption rather than creating an excess demand. 

We can refine somewhat our understanding of the effect of changes in policy by focusing on conditions in the two different regimes and asking how injection into one or the other loops will affect conditions in those regimes. 

In the investment limited regime: Increasing money flow to the labor loop results in consumer price inflation. Increasing money flow to the capital loop results in more economic growth.

In the wages limited regime: Increasing money flow to the wages loop results in more economic growth. Increasing money flow to the capital loop results in capital accumulation. 

A potential signature differentiating the two domains is a progressively higher interest rate in the investment limited regime, and a progressively lower one in the consumption limited regime. 

We see from this discussion that a concept of monetary policy as limiting inflation is more valid when monetary injection is primarily into the wages loop as was the case prior to 1980, and less consistent with current conditions when it is primarily into the capital loop. Interest rate changes can be understood to directly affect the capital loop through the relative benefits of savings over investment. On the other hand, lower interest that reduces debt burden  motivates consumers to borrow and can be a temporary help. Otherwise returning both principle and interest depends critically on increases in wages which are not present without injections into the wages loop. Indeed, consumers who borrow based upon a historical expectation that wages will increase will be disappointed of those expectations in the wages limited regime, and will remain burdened with the debt. Similarly, the current perspective that inflation and growth are linked is due to assumption of a capital limited regime.

We have considered the injection of money into either the wages or the capital loops. The point of entry however, may also be to firms that participate in both loops. It seems apparent that money given to firms promotes investment if there is investment opportunity giving rise to increases in production. One might also consider based upon rationality arguments that injection of money to firms will be generally effective because they will give it either to labor or to capital depending on what is needed for improving economic activity. However, this may not follow because the benefit to individual firms and the information that is available to them may result in a tragedy of the commons or other reasons for ineffective collective behavior. Thus it is far from apparent that they will increase wages if the best economic benefit would result from such an increase. 

\section{Conclusions}

Efforts to understand the fluctuations of economic activity are many and varied. These include not just the Goodwin model but also the Austrian business cycle theory in which temporal delays associated with credit play a central role. 

We have performed a few steps toward a better understanding of economic dynamics and the flow of money in transactions. We focused on the primary loops of wages and consumption and investment and returns, while treating the role of banks and government as providing monetary injections into those loops. Our analysis suggests that the regulation of economic activity that focuses on monetary policy has missed the problem of regulating the relative strength of flows in those two loops. The presence of two flows and one control variable can be expected on general principle not to function well. We can identify two regimes of behavior. One roughly between 1960 and 1980 in which monetary injection favored the wages and consumption loop, and one between 1980 and the financial crisis in which the injection favored the investing and returns loop. Consistent with this picture, interest rates increased until 1980 and decreased thereafter. The transition is directly consistent with tax fiscal policy changes. 

Our analysis suggests those policies---variously called Reaganomics or supply-side, free-market or trickle-down economics \cite{reagan}---which reduced taxes for the wealthy and thus promoted the availability of investment capital, occurred at a time when capital injection was needed to balance the injection of loop flows. This shift was important to avoid progressively higher inflation. The impetus for increasing flow to capital is justified before 1980 because of the net excess of demand over production, and borrowing by capital, leading to too high inflation. 

It might appear that the recession in the early 1980s is due to an imbalance between the labor and capital loop with too high a monetary flow into the labor loop. However, a more careful analysis implies that capital borrowing became strong enough so that the influx into the capital loop exceeded the need for investment in proportion to consumption, and a recession resulted. This suggests that even when flows are more directed toward the labor loop, borrowing for investment should be constrained to a level just below the needs of consumption to enable sustained growth. 

Our analysis implies that the policy changes (1) were too large so that the resulting flows were out of balance toward capital, (2) resulted in an imbalance of accumulated uninvested capital savings and consumer debt, and (3) required progressively lower interest rates that have eliminated their utility for intervention for continued growth. After 35 years, this policy has depleted the ability of the consumption loop to borrow, limiting investment so that a large excess of investment assets do not have investing opportunities.  

We can infer that the changes in flows toward capital were too high and that the balance needed resides somewhere in between. 

Consistent with our analysis of monetary flows into the labor and capital loops, a decrease in labor's share of the national income has been noted by others \cite{Elsby,Dao}. Their work does not indicate how those shares are linked to growing economic flows, or the accumulation in savings and debt.  

In the context of the limited ability of monetary policy to regulate the economy, the success of federal reserve policy to sustain economic growth over 35 years is notable. Actions taken are similar in their form to the effect of tacking a sailboat when sailing into the wind. The continued success of these effects cannot be relied upon due to exhausting the limits of interest rate reduction, and increasing reliance on alternative methods such as quantitative easing is uncertain. Monetary policy affects different aspects of the economy differently but doesn't have direct control over those differences, while fiscal policy can be more directed. Moreover, a fiscal injection balance can promote growth without monetary policy interventions. 

We recommend a shift toward policies that inject money into the wages and consumption loop to increase economic growth. We note that this same recommendation would arise from considering the problem of income inequality \cite{piketty}. Here we have shown that the effect of current income inequality is a drastic decrease in the effectiveness of economic activity and growth. 

In addition to the conclusions and summary of implications our results give rise to the following predictions:

The federal reserve will not be able to increase interest rates significantly as the trend of reduction in interest rates was not associated with the financial crisis but rather with a dynamic that started in 1980. More generally, the effectiveness of using monetary policy in and of itself as a mechanism of regulation of economic growth has reached its limits. On a time scale of a few years the next recession will take place and the dynamics of that recession will be even harder to control as the regulatory capability of monetary policy control through interest rate reduction has become essentially orthogonal to the direction of economic growth. 

The federal reserve actions in maintaining low interest rates for an extended period of time and other mechanisms of monetary injection have bought time but have not solved the problem of restoring extended economic growth. The amount of money injected into the economy by fiscal and monetary policy, including quantitative easing, by the federal reserve is large (see Fig. \ref{fig:CvsI2}). However, investment in production is not commensurate due to limited capacity for increased consumption. In simple language, a poor Labor implies there is very limited ability to increase economic activity. The recovery from recession will be short lived as the consumption of goods is not sufficient to the increase in production due to investment that is occurring. 

A new policy that redirects money for consumption is necessary for growth.

Debt relief for consumers is the primary mechanism by which increased economic activity is possible. The source of the debt relief is from savings by Capital. Traditional wealth redistribution appears to be essential for renewed economic growth. The reasons for this need are apparent in that there isn't sufficient demand and an increase in demand is needed for sustained growth. More generally, changing fiscal policy away from support for Capital (Job creators) and toward Labor (Consumers) is necessary for economic growth. This cannot be done arbitrarily but rather must be done in a measured way based upon a quantitative understanding of the economic system if we are to avoid return to the 1970s problem of high inflation.

Sustained effective economic growth will take place in economies who balance flows that are used for consumption and investment over those who favor either one. 

Our analysis is incomplete in many details and provides only preliminary steps to an analysis of  economic flows and their regulatory implications. Perhaps more importantly, it neglects the effect of international / global financial flows which are surely playing a larger role in economic activity today than in the past. 

We thank Richard Cooper, Irving Epstein, Jeff Fuhrer, Alexander Lipton,  Alfredo Morales, and Walt Vester for helpful discussions. Jean Langlois-Meurinne, Mari Kawakatsu, and Rodolfo Garcia were NECSI summer interns while students at Ecole Polytechnique, Yale University, and MIT respectively. Earlier work leading to this paper was supported by the Common Fund for Commodities. 

\appendix

\section{Money supply for economic flows \label{appendix:injection}}

Missing from the Goodwin model is the mechanism by which inflation (or growth) occurs, which is linked to changes in money supply as well as other effects. The money supply is often measured by the amounts of particular forms of money and assets (cash, bank deposits) rather than the quantities of money associated with the monetary flows we are considering here. While these may be related, the analysis of their relationship is a separate endeavor. Here we consider the money directly associated to economic flows. We clarify the mechanism of inflation or growth through injection of money into the economic flows. 

An essential observation is that while $W+R=Y$ holds at any given time $t$, the amount of transactions in each loop ($W$, $R$) and in the overall economy ($Y$) grows with inflation as well as fluctuates over time. According to the model, however, there is no source of additional money, so money must be conserved. This appears to be a paradox.

There is a conceptual way that money flows ($W$, $R$, $Y$, etc.) can be increased without increasing the amount of money in circulation---if all sectors agree to shift the money simultaneously. Consider, for example, the case where wages increase by some $\Delta W$ between time $t$ and time $t+dt$. If workers agree to increase their consumption by $\Delta C = \Delta W$ during the same time period $dt$, then the \textit{flow} of money in the labor loop effectively increases without any injection of money. This discrete transfer concept does not describe actual transactions, as it takes some time for money to transfer (i.e. to clear). The first sector to shift money must have an additional amount available to make the increased payment. Inevitably, then, the sector would have to receive or borrow money from an external source.

The flow through a loop has to increase to accommodate inflation / economic growth including the effects of population growth and changes in per capita consumption associated with changes in the types of product being produced and consumed. 

\begin{figure}[!tb]
	\centering
	\includegraphics[width=4 in]{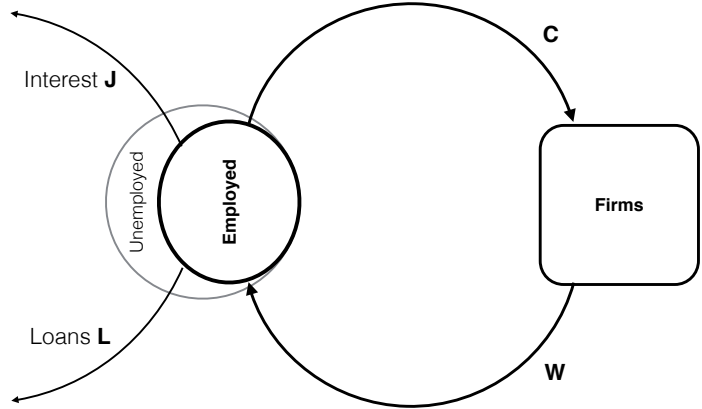}
	\caption{A schematic of the borrowing mechanism.} \label{fig:borrowing}
\end{figure}

Injection of money may arise from borrowing or fiscal injection. In principle, either workers, capital owners, or firms may all be borrowers / recipients of injection. Fig. \ref{fig:borrowing} shows the parts of the model for this analysis. 

We first describe the mechanism. We assume a slowly varying wage inflation (similar analysis holds for price inflation) summarized by a single rate $r$, and we divide each year into $N$ periods of length $\Delta t = (1/N)$ year. We begin by considering the labor loop. Workers begin the $k$th period $(1\leq k \leq N)$ with the wages from the previous period, $W_{k-1}$. During the $k$th period, they consume the amount that they have in hand plus a fraction $p_w$ of the increase in wages they expect to receive at the end of the period (similarly, for price inflation workers do without a fraction, $p_p$, of the rate of increase in consumer prices, $r_p$),
accounting for inflation and other variations. In the $k$th period, workers consume 
\begin{equation} C_k = W_{k-1}\bigg(1+\frac{p_w r}{N}\bigg) \end{equation}
To finance this consumption, workers must receive additional money, perhaps in the form of loans:
\begin{equation} L_{w_k} = W_{k-1}\bigg[\bigg(1+\frac{p_w r}{N}\bigg) - 1\bigg] = W_{k-1}\bigg(\frac{p_w r}{N}\bigg)\end{equation}
At the end of the $k$th period, they receive new wages adjusted for inflation and other variations:
\begin{equation} W_k = W_{k-1}\bigg(1+\frac{r}{N}\bigg) \end{equation}
Note that $W_k \geq C_k$. Firms receive $C_k$ but must pay $W_k$, so they must receive
\begin{equation}
L_{f_{w,k}}= W_{k-1}\bigg[\bigg(1+\frac{r}{N}\bigg) - \bigg(1+\frac{p_w r}{N}\bigg)\bigg] = W_{k-1}\bigg(\frac{(1-p_w)\ r}{N}\bigg)
\end{equation}
For the full year, workers begin the year with $W_0=W_Y/N$. Then
\begin{align}
\begin{split}
W_k &= W_0\bigg(1+\frac{r}{N}\bigg)^k \\
C_k &= W_0\bigg(1+\frac{r}{N}\bigg)^{k-1}\bigg(1 + \frac{p_w r}{N}\bigg) \\
\end{split}
\begin{split}
L_{w_k} &= W_0\bigg(1+\frac{r}{N}\bigg)^{k-1}\bigg(\frac{p_w r}{N}\bigg) \\
L_{f_{w,k}} &= W_0\bigg(1+\frac{r}{N}\bigg)^{k-1}\bigg(\frac{(1-p_w)\ r}{N}\bigg) \label{eq:labor4} 
\end{split}
\end{align}
In $N$ periods, workers accumulate a debt of
\begin{align}
\sum_{k=1}^{N} L_{w_k} & 
= W_0 \bigg(\frac{p_w r}{N}\bigg) \bigg[ \frac{ \big(1+\frac{r}{N}\big)^{N} - 1} {\big(1+\frac{r}{N}\big) - 1 } \bigg] 
= W_0 p_w \bigg[ \bigg(1+\frac{r}{N}\bigg)^{N} - 1 \bigg].
\end{align}
Similarly, firms accumulate a debt of
\begin{equation}
\sum_{k=1}^{N} L_{f_{w,k}}  = W_0 (1-p_w) \bigg[ \bigg(1+\frac{r}{N}\bigg)^{N} - 1 \bigg] \label{eq:L_w_k}.
\end{equation}
Thus, workers and firms divide the debt burden in a ratio $p_w:(1-p_w)$. Firms also have a debt related to the capital loop. Let $p_c$ be the proportion of debt in the capital loop borrowed by the capital owners. This gives rise to similar equations so that the total debt for firms is
\begin{equation}
\sum_{k=1}^{N} L_{f_{c,k}}  = R_0 (1-p_w) \bigg[ \bigg(1+\frac{r}{N}\bigg)^{N} - 1 \bigg] .
\label{eq:firmdebt}
\end{equation}
and the ratio of the debts of the three sectors is
\begin{equation}
L_{w_k}:L_{f_k}:L_{c_k}= W_0 p_w : \big[W_0(1-p_w) + R_0 (1-p_c)\big] : R_0 p_c
\end{equation}
Parameters $p_w$ and $p_c$ can be determined from these ratios and economic data.
If we consider Firm assets to be owned by Capital, borrowing of Firms and Capital are aggregated:
\begin{equation}
L_{w_k}:L_{c_k}= W_0 p_w : \big[W_0(1-p_w) + R_0]
\end{equation}

If the money flow that is linked to inflation of prices, wages, or investment is the result of borrowing, then the borrowed amount has to be repaid along with interest. If economic growth does not saturate, the original amount of the currency in circulation will eventually be dominated by the borrowed amount. Under these circumstances the borrowing has to be equal to the size of the economic activity as a whole, divided by the velocity of money. 

\section{Economic flow data \label{appendix:data}}

We classified the elements of the National Income and Product Account (NIPA) tables as described in Section V of the paper. The Bureau of Economic Analysis (BEA) adjusted the accounting to better measure GDP and match production and income measures. Many of these adjustments are not relevant to our analysis and therefore we treated them differently. The adjustments are generally small and do not affect our conclusions. We detail our treatment of reclassifications and imputations (NIPA Table 7.12) in this Appendix. 

\subsection{Return (Firm to Capital) flows}
Return flows include interest payments by firms, business current transfer payments, proprietors' income, rental income of persons, dividend payments and undistributed profits. 

In the NIPA tables (1.16.9,1.16.24), some interest payments are imputed and reclassified from bank€™ profits. This reclassification aims to include services provided without payment and reduce the discrepancy between the income and production measure of GDP. The reclassification has no impact on the total return flow which includes both interest payments and profits. The BEA adjustment assumes that financial services are furnished below their production value and implicitly paid for by customers through lower interest rates (depositor services) or higher interest rates (borrower services) in comparison with a reference interest rate (US Treasury bonds). BEA then imputes those interests to household personal income under interest receipts and personal consumption expenditures. We use the original consumption data without this imputation. 

In the NIPA tables (1.16.9,1.16.23,1.16.24), interest payments to persons are imputed from pension plans (defined benefits and defined contribution pension plans)  for holding money on their behalf. The interest payments and dividends are included in private enterprise interest and dividend payments and subtracted from undistributed profits.  As interest payments, dividend payments and undistributed profits are included in the return flow, this reclassification has no impact on our analysis.

In the NIPA tables (1.16.16), proprietors' income is adjusted to include imputations for the value of goods produced but not sold. We did not include their adjustment.

In the NIPA tables (1.16.17), rental income of persons includes an estimation of owner benefits for owning a property using an estimated rental value of the property and subtracting mortgage and consumption of fixed capital. We do not include this imputation in the return flow as it does not represent an actual cash flow.

In the NIPA tables (1.16.23), net dividends include payments to persons and government. We include only payments to persons, equivalent to NIPA Tables 2.1.15.

\subsection{Investment (Capital to Firm) flows}

In the investment flow, we include fixed investment, as well as the monetary interest receipts by firms as the total cost of investment---asset value plus borrowing costs. As described in the section on Return flows, imputed interests paid by firms to firms for services are not included because they are not actual monetary flows or borrowing costs.

In the NIPA tables (1.16.4), mortgage payments are included in interest receipts by firms. We do not include them in the calculation of investment flows as we include them in personal consumption.

\subsection{Wages (Firms to Labor) flows}

In the NIPA tables (2.1.2), wages include imputations for unpaid goods and services provided to employees (food, clothing, housing, insurance). We do not include these imputations. 

In the NIPA tables (2.1.2), wages include employer contributions pension plans. We include these even though some parts are not paid in the current period. 

\subsection{Consumption (Labor to Firms) flows}

In the NIPA tables (2.4.5U), consumption includes imputations for (a) financial services furnished without payments, (b) rental value of owner-occupied housing, and (c) value of goods produced but not sold, including (d) employer provided goods and services (food, clothing, housing). We did not include these in our analysis. 

In the NIPA tables (2.4.5U), consumption of goods and services provided by nonprofit institutions is adjusted to be the cost of providing them instead of payments. We included only the payments. 

In the NIPA tables (2.4.5U), household expenses are not included because of the inclusion of rent. We added to consumption household expenses particularly intermediate inputs (maintenance) and mortgage payments. 

\subsection{Government}

In the NIPA tables (3.1), imputations for Government depositor, borrower and insurance services furnished by banks are included. This is not included in our analysis.

In the NIPA tables (3.1), imputations for Government underfunding of pension plans is considered as a loan from firms to government for which an imputed interest to government is included. Government savings are reduced. These changes are not included in our analysis.

We estimated Labor and Capital to Government flows by separating income taxes between them. We assume that in 2014 Capital paid $37.9\%$ of total income taxes as reported by the IRS for the top $1\%$ of earners making more than $\$465,000$ in adjusted gross income \cite{IRS}. We then calculate the fraction of historical taxes from the prevailing maximum tax rate using 
\begin{align}
T&=T_C+T_L =S_C T+S_L T \\
T_C&=R_C I_C = S_CT\\
S_C&=R_C  (I_C/W)(W/T)
\end{align}
where $T$ is the total income tax, $T_C$ and $T_L$ are the taxes paid by Capital and Labor respectively accounting for fraction $S_C$ and $S_L$ respectively, $I_C$ is the income of Capital and W is the income of Labor. $R_C$ is the tax rate paid by capital. To calculate the final expression we substitute for $R_C$ the maximum tax rate, the second factor is assumed to be fixed over time to the value in 2014 and calculated given the value of $S_C$, and the final factor is available from NIPA data. Due to the estimates used, $S_C$ may be greater than $100\%$, and we threshold it at that value. Alternative assumptions for the value in 2014 for the fraction of taxes paid by Capital ranging from $20\%$ to $45\%$ give the same conclusions. 

\subsection{International}

Our analysis does not include international interactions. Corrections due to global flows do not affect the conclusions. Because of globalization, international trade has increased since 1990, and these interactions should be considered in further improvements of the model.

\end{document}